\documentclass[fleqn,12pt,twoside]{article}
\usepackage{espcrc1}
\newcommand{\AmS}{{\protect\the\textfont2
  A\kern-.1667em\lower.5ex\hbox{M}\kern-.125emS}}

\usepackage{natbib}

\def\sep{;~~}

\usepackage{amssymb}
\usepackage{amsfonts}
\newcommand{\text}{\mbox}
\def\Q{{\mathbb{Q}}}
\def\H{{\mathbb{H}}}
\def\R{{\mathbb{R}}}
\def\N{{\mathbb{N}}}
\def\C{{\mathbb{C}}}

\def\OR{\vee}

\def\union{\cup}

\def\fsl{\mathfrak{sl}}
\def\fsu{\mathfrak{su}}

\def\openone{\leavevmode\hbox{\small1\kern-3.8pt\normalsize1}}
\def\RR{{\rm I\kern-.2emR}}
\def\tr{{\rm tr}\; }
\def\ce{{\mathcal E}}

\def\cd{{\mathcal D}}
\def\cb{{\mathcal B}}

\def\ch{{\mathcal H}}
\def\cs{{\mathcal S}}
\def\ct{{\mathcal T}}
\def\cp{{\mathcal P}}
\def\cm{{\mathcal M}}
\def\cf{{\mathcal F}}
\def\ci{{\mathcal I}}
\def\ca{{\mathcal A}}
\def\co{{\mathcal O}}
\def\calr{{\mathcal R}}
\def\cv{{\mathcal V}}
\def\cz{{\mathcal Z}}
\def\cx{{\mathcal X}}

\newcommand{\epigram}[3]
{
\begin{flushright}
\baselineskip=13pt
\parbox{4.2in}{\baselineskip=13pt
``#1''}\medskip\\
---{\it #2}\\
#3
\end{flushright}
}

\newcommand{\ket}[1]{| #1 \rangle}
\newcommand{\bra}[1]{\langle #1 |}

\newcommand{\outerp}[2]{\ket{#1}\! \bra{#2}}

\newcommand{\projj}[2]{\ket{#1}\ket{#2}\! \bra{#2}\bra{#1}}
\newcommand{\kett}[2]{\ket{#1}\ket{#2}}
\newcommand{\braa}[2]{\bra{#1}\bra{#2}}

\newcommand{\beq}{\begin{equation}}
\newcommand{\eeq}{\end{equation}}
\newcommand{\beqa}{\begin{eqnarray}}
\newcommand{\eeqa}{\end{eqnarray}}
\newtheorem{definition}{Definition}

\newtheorem{problem}{Problem}
\newtheorem{theorem}{Theorem}
\newtheorem{proposition}{Proposition}

\newtheorem{conjecture}{Conjecture}

\def\QED{\mbox{\rule[0pt]{1.5ex}{1.5ex}}}

\title{Quantum information processing, operational quantum logic, 
convexity, and the foundations of physics \\
\vskip 8pt
Los Alamos Technical Report LA-UR 03-1199}

\author{Howard Barnum
\thanks{Thanks to the US DOE for financial support.}
\address[LANL]{CCS-3: Modelling, Algorithms, and Informatics. Mail Stop B256 \\
Los Alamos National Laboratory, Los Alamos, NM 87545
USA.  {\tt barnum@lanl.gov}}}

\begin{document}
\maketitle

\begin{abstract}

Quantum information science is a
source of task-related axioms whose consequences can be explored
in general settings
encompassing quantum mechanics, classical theory, and more.
Quantum states are
compendia of probabilities for the outcomes of possible
operations we may perform a system: ``operational states.''  I
discuss general frameworks for ``operational theories''
(sets of possible operational states
of a system), in
which convexity plays key role.  
The main technical content of the paper is in a theorem that any
such theory
naturally gives rise to a ``weak effect algebra'' when outcomes having 
the same probability in all states are identified,  
and in the introduction of a notion of ``operation algebra'' that 
also takes account of sequential and conditional operations.
Such frameworks are 
appropriate
for investigating what things look like from an ``inside
view,'' {\em i.e.} for describing perspectival information that one
subsystem of the world can have about another.  Understanding how such
views can combine, and whether an overall ``geometric'' picture
(``outside view'') coordinating them all can be had, even if this
picture is very different in structure from the
perspectives within it, is the key to whether we may be
able to achieve a unified, ``objective'' physical view in which
quantum mechanics is the appropriate description for certain
perspectives, or whether quantum mechanics is truly telling us we must
go beyond this ``geometric'' conception of physics.
\end{abstract}

{\em Keywords:} quantum information \sep foundations of quantum mechanics \sep
quantum computation \sep quantum logic \sep convexity
\sep operational theories
PACS codes: 





\section{Introduction}

The central question quantum mechanics raises for the foundations of
physics is whether the attempt to get a {\em physical} picture, from
``outside'' the observer, of the observer's interaction with the
world, a picture which views the observer as part of a reality which
is at least roughly described by some mathematical structure, which is
interpreted by pointing out where in this structure we, the observers
and experimenters, show up, and why things end up looking as they do
to observers in our position, is doomed.  The ``relative state''
picture that arises when one tries to describe the whole shebang by an
objectively existing quantum state is unattractive, and many seek to
interpret quantum states instead as subjective, ``information'' about
how our manipulations of the world could turn out.  Whatever else they
may be the quantum states of systems clearly {\em are} compendia of
probabilities for the outcomes of possible operations we may perform
on the systems: ``operational states.''  
An {\em operational theory}
is a 
specification of the set of possible operations on a system
and a set of admissible operational states.    
This ``operational'' point
of view can be useful whether one wants to
consider the operational theory as for some reason all we can hope for, 
{\em or} as a description
of how perspectives look within an overarching theory such as the
relative state interpretation (RSI).  
While it
has not yet made a decisive contribution toward
resolving this tension, by focussing on the role of information held
(through entanglement or correlation) or obtained (by measurement) by
one system about another QIP concentrates one's attention on the
practical importance of such measurements, and develops flexibility in
moving between the inside and outside views of such
information-gathering processes.  It thus provides tools and concepts,
as well as the ever-present awareness, likely to be useful in
resolving this tension, if that is possible.  

This paper is dedicated to the memories of two researchers in
quantum foundations, who I knew only through their collaborators and
their work: Rob Clifton and Gottfried T. (``Freddy'') R\"{u}ttiman.
They will continue to influence and inspire for the duration of the
intellectual adventure of understanding, at the deepest level, our
theories of the world.  Their work is particularly relevant to the
themes of this paper.  Algebraic quantum field theory is an example of
integrating local perspectives (local $^*$-algebras of observables)
into a coherent overall structure; 
Clifton made deep investigations into foundational issues in 
AQFT---for example, \cite{CH2001a} considers
entanglement in this setting.  
He was also involved in one of the most spectacular
successes to date of the project of applying quantum
information-theoretic axioms to quantum foundations \citep{CBH2002a}.
R\"{u}ttiman's work involved, for example, linearization theorems for
lattice-based quantum logics \citep{Ruttiman93a} which parallel and
prefigure the ones discussed herein for convex effect-algebras, and
investigation of the relation between the property lattice and face
lattice of a state space \citep{Ruttiman81a}.

The paper is organized as follows.
Section \ref{sec: peculiar} considers
some salient general implications of QIS for foundational
questions (irrespective of its contributions to this project).
Section \ref{sec: RSI vs subjective}
discusses the  
relative state and ``subjective'' views on the
foundations of quantum mechanics.
Section \ref{sec: combination of perspectives} discusses whether
and how the perspectives of different observers can be combined,
via tensor products and other constructions.
Section \ref{sec: operational theories}
constructs ``weak effect algebras''
from probability compendia via identification of probabilistically 
equivalent outcomes, 
reviews operational quantum logic, 
especially convex effect algebras, 
and introduces the notion of operation algebra which formalizes 
the notion of doing operations in sequence, possibly conditioned on 
the results of previous operations.  
In Section \ref{sec: applying operational logic}, I  briefly consider
uses of the framework in applying QIP ideas to
foundational questions.

A major part of empirical 
quantum logic is ``deriving quantum mechanics.''  The hope is that
if this can be done with axioms whose operational, information-processing,
or information-theoretic meaning is clear,
then one will have a particularly nice kind of answer to the question
``Why quantum mechanics?''  
QI/QC provides a source of axioms, with
natural interpretations involving the possibility or impossibility of
information-theoretic tasks.  This is likely to
contribute to {\em whichever} mode of resolution turns out to be
right.  Within the ``geometric'' or ``objective overall picture''
resolution, one might obtain the answer: Why quantum mechanics?
``Because it's the sort of structure you'd expect for describing 
certain perspectives (of the sort beings
like us wind up with) that occur ``from the inside point of view'' within an
overarching picture of {\em this} [fill in the blank] sort.''  The
blank might be filled in with a specific overarching physical
theory, or with fairly general
features.  A
similar answer might arise from the more ``subjectivist'' point
of view on quantum states.  Why quantum mechanics?  ``Because it's the
sort of structure you'd expect for describing the perspectives
``from the inside point of
view'' within a reality of {\em this} sort, which
reality is however not completely describable in physical terms, so
that these perspectives are as good as physics ever gets.''
Those who anticipate or hope for a physical
picture, including relative state-ers, and
those who think such an overarching {\em physical} picture unlikely to
emerge, can nevertheless fruitfully pursue similar projects 
using axiomatic arguments involving the notion of
``operational theory'' to derive quantum mechanics, to understand,
how it differs from or is similar to other conceivable
theories, and the extent to which it
does or does not follow from elementary conceptual
requirements (one way in which it could be ``a law of
thought'') or, in a more Kantian or perhaps ``anthropic'' way, from
the possibility of rational beings like us (a different way in which
it could be ``a law of thought'').  
Details might depend on one's orientation: 
subjectivists might be more
inclined to axioms stressing the formal analogies between density
matrices and probability distributions, and between quantum ``collapse'' and
Bayesian updating of probability distributions \citep{Fuchs2001a}.
But since on the ``overarching physical picture with
perspectives'' view the probabilities are also tied to a
``subjective,'' perspectival element, the Bayesian analogy is
quite natural on this picture too.
The close link between ``empirical
operational theories'' and perspectival information that one subsystem
of the world can have about another, and the importance of {\em tasks}, of
what can and cannot be done from a given perspective, 
suggests that generalized information theory and 
information processing, of which QIS supplies a main example,
will play a major role in this project.

\section{QIP: The power of the peculiar}
\label{sec: peculiar}
Virtually all of the main aspects of quantum mechanics exploited in
QIP protocols have been understood for
decades to be important peculiarities of quantum mechanics.
The
nonlocal correlations allowed by entanglement are
exploited by better-than-classical
communication complexity protocols \citep{Buhrman97a}; the necessity
of disturbance when information is gathered on a genuinely quantum
ensemble \citep{Fuchs95b, Barnum98d, Barnum2001a, Banaszek2001a,
Bennett94a, Barnum2001b}, closely related to the ``no-cloning
theorem'' \citep{Wootters82a} and no-broadcasting
theorem \citep{Barnum96a, Lindblad99a}, is the basis of quantum cryptography;
the ability to obtain information complementary to that available in
the standard computational basis
is the heart of the historic series of algorithms due to
\citet{Deutsch85a}, \cite*{Deutsch92a},
\citet{Simon97a},
\cite*{Bernstein97a}, and culminating in 
Shor's (\citeyear{Shor94a,Shor97a})
polynomial-time factoring algorithm.
These peculiarities
are no longer just
curiosities, paradoxes, philosophers' conundrums, they now have
worldly power.

A number of more specific and/or technical points on which QIP has 
contributed, or shows potential to contribute, something
new to old debates can be identified.  
First, QIP provides tools with which to analyze much more precisely and
algorithmically questions of what can and cannot be measured
\citep{Wigner52a,Araki60a,Reck94a}, or
otherwised accomplished, either precisely or approximately, in quantum
mechanics.  
Some measurements are even uncomputable in essentially the same
sense as are some partial recursive functions in classical computer science.
This raises the issue of the extent to which ``operational''
limitations, including basic and highly theoretical ones such as
computability, should be built into our basic formalisms, and what it
means for the interpretation of those formalisms and the ``reality''
of the objects they refer to, if they are not.
Second, QIP techniques and concepts 
such as 
error-correction and active and passive stabilization and
control 
promise to allow a much more systematic
approach than previously 
to experiments and thought-experiments suggested by foundational
investigations.  
Third, QIP
has
demonstrated the power of taking the formal analogy between quantum
density matrices and classical probability distributions seriously.
Most things one does with probability distributions in classical
information theory 
have (sometimes
multiple) natural
quantum analogues when quantum states replace probability
distributions.  
Fourth, QIP 
provides a source of natural ``operational'' questions about
whether certain information-processing tasks can or cannot be
performed, usable when considering empirical theories more general
than quantum mechanics.
Also, QIP may be a natural
source of examples of empirical theories.  These
arise when one considers attempts to perform quantum information 
processing with the restricted means
available in some proposed implementation
of quantum computing.
For example, QIP considerations stimulated some of us \citep{Barnum2002a}
to generalize
the notion of ``entanglement'' to pairs of lie algebras and
beyond that to pairs of ordered linear spaces.
\section{Relative state vs. information interpretation of quantum mechanics}
\label{sec: RSI vs subjective}

The central tension in interpreting quantum mechanics is between the
idea that we are part of a quantum world, made of quantum stuff
interacting with quantum stuff, evolving according to the
Schr{\"o}dinger equation, and the apparent fact that when we evolve so
as to correlate our state with that of some other quantum system
which is initially in a superposition, we get a single measurement
outcome, with probabilities given by the squared moduli of coefficients
of the projections of the state onto subspaces in which we see a 
definite measurement outcome.  The RSI 
reconciles these ideas by taking the view that the experience of
obtaining a definite measurement result is how things appear
from one point of view, our subspace of the world's Hilbert space, and
the full state of the world is indeed a superposition.  As I see it
the correct way, on this view, to account for the appearance that
there is a single measurement result, is the idea that the experience
of a conscious history is associated with definite measurement
results, so that consciousness forks when a quantum measurement is
made \citep{Barnum90a}.  
Just as there is no consciousness whose experience is that of
the spacetime region occupied by you, me, Halley's comet, and the left
half of Georges Sand, so, after a measurement has correlated me with
the the z-spin of an initially x-polarized photon, there is no
consciousness whose experience is that of the full superposition (or,
once these branches of me are decohered, of the corresponding
mixture).  
Understanding why this happens as it does 
would appear to involve 
psychological/philosophical considerations about how minds are individuated.
A more precise account must await a
better scientific understanding of consciousness, though there are
probably some useful things to be said by philosophers, psychologists,
biologists, and decoherence theorists.  It is deeply bound up with the
problem of choosing a ``preferred basis'' in the relative state
interpretation (i.e., the question, ``relative to what?''), and also
with the problem of what tensor factorization of Hilbert space to
choose in relativizing states, which appears in this light as the
question of which subsystems of
the universe support consciousness.  The stability of
phenomena and their relations enforced by decoherence may underly the
ability to support consciousness.

Despite 
sometimes conceding when pressed that they can't show the RSI is inconsistent,
its opponents also sometimes claim it is inconsistent for an observer to
view him or herself as described by quantum theory \citep{Fuchs2000a}.
I am not aware of a rigorous argument for this, though.  Even an 
argument within a toy model would be valuable. 
But ven if it is shown that it would be
inconsistent {\em for an observer herself} to have a complete
quantum-mechanical description of herself, the system she is
measuring, and the part of the universe that decoheres her ``in the
pointer basis,'' that does not show that such a description
is itself inconsistent.  
Similar ``bizarre self-referential
logical paradoxes'' \cite{Fuchs2000a} seem just as threatening (or not)
for a classical description.

Some \cite{Bilodeau96a} think that QM is telling us we must abandon
the ``geometric'' conception of physics as giving us an ``outside view''
of reality. 
But I think that 
rather than just welcoming the ability to view quantum mechanics
as only appropriate to describing an observer's perspective on a
system, revelling in the subjectivity of it all, the way it perhaps
leaves room for mind, freewill, etc... as unanalyzed primitives,
it is still promising to try to get a grip on these matters
``from an outside point of view.''  
An analogy might be special relativity.  Here, an overarching picture
was achieved by taking seriously the fact that position and time
measurements are done via operations, from the perspective of
particular observers.  The heart of the theory is to coordinate those
perspectives into a global Minkowski space structure, explaining in the
process certain aspects
of the local operational picture (like restrictions on the values of
velocity measurements).
I don't think that we should yet give up
on an attempt at such coordination in the quantum case, perhaps celebrating the
supposed fact that quantum mechanics has shown us that it will be impossible
to achieve under the aegis of physics.

An important point brought out by the attempt at a relative state
interpretation of quantum mechanics is the need to bring in, in
addition to Hilbert space, notions of preferred subsystems
(``experimenter'' and ``system'' perhaps also the ``rest of the
world'') or preferred orthogonal subspace decompositions (choice of
``pointer basis'' \citep{Zurek81a}).  It seems unlikely, as Benjamin
Schumacher likes to point out, that a Hilbert space, Hamiltonian, 
and initial state,  
will single out preferred subspace decompositions in which dynamics looks
nontrivial, hence the
RSI should involve aspects
of physics beyond Hilbert space.  
Schumacher also
points out that a Hamiltonian evolution on a Hilbert space can be made
to look trivial by a time-dependent change of basis.  If one takes the
view that ``the classical world'' is supposed to {\em emerge} from
this structure (Hilbert space, Hamiltonian, and initial state), then
perhaps such transformations are legitimate.  On the other hand, they
are not wholly trivial: if one specifies a Hamiltonian
dynamics on a Hilbert
space, one is implicitly specifying two groups of canonical isomorphisms
between a continuum of Hilbert spaces, 
continuously parametrized by time.   One of them says what we mean by 
``same Hilbert space at different times,'' providing a framework with 
respect to which 
we can then define a Hamiltonian evolution specified by the other
one.
If we could pick out a set of subspaces that
are special with respect to this structure, that would be interesting.
I have doubts that we can; I also like Schumacher's criticism that
this specification of ``two connections on a fiber bundle instead of
just one'' seems mathematically unnatural.  But I am not wholly
convinced by Schumacher's criticisms.  I view the RSI less as a way of
getting the classical world emerge from Hilbert space, and more as a
way of giving a realistic interpretation to Hilbert space structure in
the presence of additional structure such as preferred bases or
subsystem decompositions that represent other aspects of physics.
Schumacher views his arguments as showing that one
needs these additional aspects of physics---''handles on Hilbert
space''---to get a canonical identification of, say, bases from one
time to the next (say the spin-up/down basis in a given reference frame).  
He interprets
this as showing the appropriateness of Hilbert space descriptions
for subsystems where the special structure lies in
relations to other systems (such as measuring appartus), and the
inappropriateness of the Hilbert space structure for the description
of the whole universe.  There are plenty of such
non-Hilbert space aspects of physics, involving symmetries, spacetime
structure.  The need for renormalization and the difficulties with
quantum gravity suggest some difficulty in squaring quantum mechanics
with some of these ``geometrical,'' ``outside'' aspects of physics.
Perhaps the distasteful aspects of the quantum-mechanical outside view
may vanish once such a squaring, with whatever flexing is necessary
from both sides, is accomplished.

Bell showed that nonlocal hidden variables
are the only non-conspiratorial way to realistically
model the statistics of quantum measurements.  (Non-conspiratorial
refers to a prohibition
on explaining the statistics of quantum
measurements by correlations between the hidden variables and what
we ``choose'' to measure.)  But when we are
contemplating quantizing the spacetime metric or otherwise 
unifying gravity and quantum mechanics, perhaps it is not
too farfetched to  imagine that spacetime and causality will
turn out to be emergent from a theory describing a structure at
a much deeper level....if this structure contains things
whose effects, at the emergent level of spacetime, can be interpreted
as those of ``nonlocal hidden variables,'' this should
hardly surprise or dismay us.  

My view toward the RSI with macroscopic superpositions
is much like Einstein's toward taking quantum mechanics as a complete
physical theory:  I just don't think
the universe is like that.  
\cite{Schulman97a} proposes to retain essentially a one-Hilbert space,
state-vector evolving according to the Schrodinger equation,
no-collapse version of quantum mechanics, interpreted realistically, 
but to 
bring in cosmology and statistical mechanics and argue that
symmetric consideration of final conditions along with the usual
initial conditions (that the universe was once much
denser and hotter than it is now) rules out macroscopic
superpositions.  There is a lot to do to make this persuasive.
It is certainly an ingenious and appealing
idea.  And if it does work, I am fairly happy to retain the rest of
the relative state metaphysics, now that I will not be committed to
the disturbing existence of forking D\"oppelgangers in subspaces of
Hilbert space decohered from me.

\section{The combination of perspectives}
\label{sec: combination of perspectives}
We should continue to
investigate both the inside and outside views of quantum systems, and in
interpretational matters to pursue a better understanding both of the
possibility of viewing quantum theory as about the dynamics of
information-like, perhaps subjective, states, and of the possibility
of viewing it as about the sorts of entanglement and correlation
relations that can arise between systems.  A prime example of a
worthwhile program along the former lines is the
Caves-Fuchs-Schack Bayesian approach;  a
prime example of a worthwhile program along the latter lines is
understanding how the probabilities for collapse can be understood
within the RSI \cite{Deutsch99a, Wallace2002a}, 
also as something like a 
Bayesian process of ``gaining more information about which
branch of the wavefunction we are in.'' 
The 
similarity between these two programs is an example of how
the operational approach is relevant to both:  investigate quantum
mechanics' properties as a theory of perspectives of subsystems
on other systems, without prejudging whether or not the perspectives
will turn out to be coordinatable into an overarching picture---indeed,
while trying to ferret out how this might happen or be shown to be
inconsistent, and how this possibility or impossibility may be
reflected in the operational, perspective-bound structures.

The 
Rovelli-Smolin ``relational quantum mechanics'' 
approach 
suggests ways in which 
quantum mechanics could be good for describing things from the point 
of view of subsystems, but not appropriate for the entire universe, 
but in which {\em nevertheless} there exists a mathematical 
structure---something like a topological quantum 
field theory (TQFT) or spin foam---in 
which these local subsytem points of view are coordinated
into an overall mathematical structure which, while its terms may be
radically different from those we are used to, may still be viewed
as in some sense ``objective.''  It is still far from clear
that this can allow us to avoid the more grotesque aspects
(proliferating macroscopic superpositions viewed as objectively
existing) and remaining conceptual issues (how to identify a preferred
tensor factorization, and/or preferred bases, in which to identify
``relative states'') of the Everett interpretation. 
 
In TQFT's or spin networks and generalizations, 
the description appropriate to ``perspectives'' is still
Hilbert spaces, but only in special cases do these combine as tensor
products.  If we view a manifold as divided into ``system'' and
``observer'' via a cobordism, then as the ``observer'' gets small
enough, while the ``system'' gets larger, we start getting, not the
increase in Hilbert-space size to describe the system that we might
expect as the system gets larger, but a decrease in Hilbert-space size
whose heuristic interpretation might be that the observer has gotten
so small that it no longer has the possibility of measuring all the
operators needed to describe the ``large'' Hilbert space one might
have expected.  The Hilbert space does not describe the ``large'' rest
of the world; it describes the relation between a small observer and
the larger rest of the world.
In these theories,  we might see how the quantum description
of certain perspectives could arise as a limiting case of some more
general type of perspective, which necessarily also arises in an
overarching structure that includes quantum-mechanical perspectives in
a physically reasonable way.  Or we might see how a non-tensor product
law of combination of subsystems---quantum or not---could be relevant
in some situations.  This is just the sort of thing that operational
quantum logic aspires to investigate, and that might be related to 
the ability to perform, or not, information-processing tasks.  

\section{Frameworks for empirical operational theories}
\label{sec: operational theories}
In this section I will introduce frameworks 
I find particularly useful for thinking
about empirical operational theories.  
David \cite{Foulis98a} has provided a good review of the general area of 
mathematical descriptions of operational theories (which he
calls ``mathematical
metascience'').  That review stresses 
concepts similar to those I use here, 
notably that of 
effect algebras,'' introduced under
this name by Foulis and Bennett \citep{Foulis94a}, but also,
as ``weak orthoalgebras'' in
\cite{Giuntini89a}, and independently, in an order-theoretic
formulation, as ``difference posets'' (D-posets, for short) by
\cite{Kopka94a}.  Longer and more technical introductions
are available in \cite{Foulis2000a} and \cite{Wilce2000a}.

\subsection{Probabilistic equivalence}
My preferred approach to operational theories starts from
the compendia of probabilities, that 
are empirically found to be 
possible for the different results of different possible
operations on a system, and constructs various more abstract
structures for representing aspects of empirical theories---effect
algebras, classical probability event-spaces, $C^*$-algebraic
representations, spaces of density operators on Hilbert spaces,
orthomodular lattices, or what have you---from these.  
With most such types of abstract
structures, the possibility of constructing them 
from phenomenological theories (sets of compendia of probabilities
for measurement outcomes) will impose
restrictions on these sets of compendia, and the nature of
these restrictions constitutes the empirical significance of the
statement that our empirical theory has this abstract
structure.  This approach promises to systematize our understanding
of a wide range of empirical structures and their relationships, both
mathematically and in their empirical significance.  
The relationship to the
probabilities of experimental outcomes has always been a critical part
of understanding these structures as empirical theories. 
The space of ``states'' on such
structures is also often a crucial aid to understanding their abstract
mathematical
structure.  This is of a piece with the situation in many categories
of mathematical objects.  $[0,1]$ is a particularly simple example of
many categories of ``empirical structure,'' and a state is a morphism
onto it; understanding the structure of some more complex object in
the category in terms of the set of all morphisms onto this simple
object is similar to, say, understanding the structure of a group in
terms of its characters (morphisms to a particularly simple group).

In this project I make use of an idea which has come in for a
fair amount of criticism, but has been with us from early in the game
(cf. e.g. \citep{Mackey63a}, \cite{Cooke81a}
(who even ascribe it to Bohr), 
\cite{Ludwig83a}, \cite{Mielnik69a} p. 14).
This is the notion of ``probabilistic equivalence'': two outcomes, of
different operational procedures, are viewed as equivalent,
if they have the same probability ``no matter how the
system is prepared,'' i.e., in all admissible states of the
phenomenological theory.  An interpretation of equivalent outcomes
as ``exhibiting the same
effect of system on apparatus'' is probably due to Ludwig, perhaps
motivating his term ``effect''
for these equivalence classes (at least in the quantum case).  
It helps forestall the objection that two outcomes equivalent
in this sense may lead to different probabilities
(conditional on the outcomes) for the results of further
measurements.  They are equivalent only as concerns
the effect of the system on the apparatus and observer, not vice
versa.  The criticism implicitly supposes a framework in which
operations may be performed one after the other, so that outcomes of
such a sequence of $N$ measurements are strings of outcomes $a_1 a_2
... a_N$ of individual measurements.  Then a stricter notion of
probabilistic equivalence may be introduced, according to which two
outcomes $x$ and $y$ are equivalent if for every outcome $a,b$ 
the probability of $axb$ is the same as
that of $ayb$, in every state.  

Before considering in detail the derivation of the structure of the
set of probabilistic equivalence classes (``effects'') of an
operational theory, I will introduce some of the abstract structures
we will end up with: effect algebras and ``weak effect algebras,''
motivating them (in the case of effect algebras) with classical and
quantum examples.

\begin{definition}
An {\rm effect algebra} 
is an object $\langle \ce, 1, \oplus \rangle$, where
$\ce$ is a set of ``effects,'' $1 \in \ce$, 
and $\oplus$ is a partial binary operation on $\ce$ which 
is (EA1) strongly commutative and (EA2) strongly associative.  
The qualifier ``strongly,'' which is not redundant only because
$\oplus$ is partial, indicates that if the sums on
one side of the equations for commutativity and associativity
exist, so do those on the other side, and they are equal.
In addition,
(EA3) $\forall e \in \ce, \exists ! f \in \ce ~~(e \oplus f = u)$.
(The exclamation point indicates uniqueness.
We give this unique $f$ the name $e'$; it is also called the
{\em orthosupplement} of $e$.)
(EA4) $a \oplus 1$ is defined only for $a = 1'.$  (We will often call
$1'$ by the name ``0''.)
\end{definition}
If we only require that the equalities specifying associativity
($a \oplus (b \oplus c) = (a \oplus b) \oplus c$)
and commutativity ($a \oplus b = b \oplus a$)
hold when both sides are
defined, allowing the possibility that 
one is defined while the other is 
not, we call these ``weak commutativity'' and ``weak associativity.''

In the effect algebra $\ce(\ch)$ of quantum mechanics (on a finite-dimensional 
Hilbert space $\ch$, say), $\ce$ is the 
unit interval of operators $e$ such that 
$0 \le e \le I$ on the Hilbert space,
$\oplus$ is 
ordinary addition of operators restricted to this interval
(thus $e \oplus f$ is 
undefined when $e + f > I$), $1$ is the
identity operator $I$, and $e' = I - e$, so $0$ 
is the zero operator.
A classical example is the set $\cf$ of 
``fuzzy sets'' on a finite
set $\Lambda = \{\lambda_1,...,\lambda_d\}$ (which are 
functions from $\Lambda$ to $[0,1]$), with  $\oplus$ 
as ordinary pointwise addition of functions (i.e. defining 
$f+g$ by $(f+g)(x) = f(x) + g(x)$ except that $f \oplus g$ is 
undefined when $f + g$'s range is not contained in $[0,1]$), 
and $1$ the constant function whose value is $1$.
$\langle \cf, 1, \oplus\rangle$ is an effect algebra obviously
isomorphic to the restriction of the quantum effect algebra on a
$d$-dimensional Hilbert space to effects which are all diagonalizable
in the same basis.  These ``fuzzy sets'' may be interpreted as the
outcomes of ``fuzzy measurements'' in a situation where there are
$d$ underlying potential atomic ``sharp'' measurement results or
``finegrained outcomes,'' but our apparatus may have arbitrarily 
many possible meter readings, connected to these ``atomic outcomes''
by a noisy channel (stochastic matrix of transition probabilities, 
which are in fact the $d$ values taken 
by the function (effect) representing a (not necessarily atomic)
``outcome''.).

We consider various modifications of the effect algebra notion.
We introduce ``weak effect algebras'' 
which are EA's in which
strong associativity (EA2) is replaced by weak associativity.
An {\em orthoalgebra} instead
adds the axiom (OA5) that $x \oplus x$ exists only for $x=0$.  
The projectors on a quantum-mechanical system, with the same
definitions of $1, \oplus$ as apply to more general POVM elements,
are an example 
(as well as being a sub-effect algebra of
$\ce(\ch)$).  
Wilce considered ``partial abelian
semigroups,''  (PASes) which require 
only (EA1) and (EA2); various combinations of additional
requirements then give a remarkably wide variety of algebraic structures
that have been considered in operational quantum logic, including effect
algebras, test spaces, E-test spaces, and other things.
In particular, an effect algebra is a positive, unital, cancellative,
PAS (see below).  

A {\em state} $\omega$  on a weak effect algebra $\langle \ce, \oplus, 
 1 \rangle$
is a function from $\ce$ to $[0,1]$ satisfying:
$
\omega(a \oplus b) = \omega(a) + \omega(b)\;,
\omega(1) = 1\;.
$
A {\em finite resolution of unity} in a weak effect algebra (to be interpreted
as the abstract analogue of a measurement) is a set $R$ 
such that $\oplus_{a \in R}  a = 1$.   
So for a resolution of unity $R$, 
$\sum_{a \in R} \omega(a) = 1$: the probabilities of measurement results
add to one.   
A {\em morphism} from one WEA $\ce$ to another $\cf$
is a function $\phi: \ce \rightarrow \cf$ such 
that $\phi(a \oplus b) = \phi(a) \oplus \phi(b)$;  it is called 
{\em faithful} if in addition, $\phi(1_\ce) = 1_\cf$, where $1_\ce$ and
$1_\cf$ are the units of $\ce$ and $\cf$.   $[0,1]$, with $\oplus$ addition
restricted to the interval, is an effect algebra, so a state on $\ce$ 
is a faithful morphism from $\ce$.

I will attempt to avoid issues involving effect algebras and WEA's
where $\ce$ is infinite and infinite resolutions of unity are defined,
though finite dimensional quantum mechanics is properly done that
way.  
(\citep{Feldman93a},
\citet{Bugajski2000a} and \citet{Gudder2000a}, for example treat 
these issues.)
To this end I will assume that EA's and WEA's are {\em locally finite}:
resolutions of unity in them have finite cardinality.
For finite $d$-dimensional quantum mechanics, most
things should work the same if we restrict ourselves to work with
resolutions of unity into $d^2$ elements.  

Now, I will relate this abstract structure to phenomenological theories, 
by showing that 
one can derive a natural weak effect algebra from any phenomenological
theory.
The operation $\oplus$ of the weak effect algebra will be the image,
under our construction, of the binary relations OR ($\OR$) in 
the standard propositional logics (one for each measurement)
of propositions about the outcomes 
of a given measurement.   (This is one justification for calling effect
algebras ``logics''.) 

In order to describe this construction, we first review Boolean algebras.
A Boolean algebra is an orthocomplemented distributive
lattice.  A lattice is a structure $\langle L, \vee, \wedge \rangle$,
where $L$ is a set, $\vee$,$\wedge$ total
binary operations on $L$ with the
following properties.
Both operations are associative, commutative, and idempotent
(idempotent means, e.g., $(a \wedge a = a)$).  In
addition, together they are {\em absorptive}:
$a \wedge (a \vee b) = a\;, \nonumber 
a \vee (a \wedge b) = a.$
$\vee$ is usually called {\em join}, $\wedge$ is usually called
{\em meet}.  
These properties are satisfied by letting $L$ be any powerset (the 
set of subsets of a given set), and the operations $\vee, \wedge$ 
correspond to $\cup, \cap$.  
For $L= 2^X$ (the power set of $X$) we call this lattice the 
{\em subset lattice} of $X$.
An important alternative characterization of a lattice is as a 
set partially ordered by a relation we will call $\le$.  
If every pair $(x,y)$ of elements have both a greatest lower bound
(inf) and a least upper bound (sup) according to this ordering, 
we call these
$x \wedge y$ and $x \vee y$, respectively, and the set is a lattice
with respect to these operations.  Also, for any lattice as defined
above, we may define a partial ordering $\le$ such that $\wedge$,
$\vee$ are inf, sup, respectively, in the ordering.  So the two
characterizations are equivalent.

A lattice is said to be {\em distributive} if meet distributes over 
join:
$a \vee ( b \wedge c) = (a \vee b) \wedge ( a \vee c) \;.
$
(This statement is equivalent to its dual (the statement with
$\wedge \leftrightarrow \vee$).)
If $L$ contains top and bottom elements with respect to $\le$, we
call them $1$ and $0$.  They may be equivalently be defined via
$a = a \wedge 1,$ $a = a \vee 0$ for all $a \in L$.
We define $b$ to be a {\em complement} of $a$ if $a \wedge b = 0$
and $a \vee b$ = 1.  
Complements are unique in distributive lattices, not necessarily so
in more general lattices.  When all complements are unique, we write
complementation as a unary relation (operation) $'$;  this relation is
not necessarily total even in distributive lattices with $0,1$.
A {\em Boolean lattice}, or Boolean algebra, is a distributive lattice
with $0,1$, in which every element has a complement.  
Any subset lattice $L=2^X$ is a Boolean algebra, with $0 = \emptyset$
and $1 = X$.  
\begin{definition}
A (locally finite) 
phenomenological theory $\cp$ is a set $\cm$ of disjoint finite sets
$M$, together with a set $\Omega$ of functions (``states'') $\omega$
from (all of) $\union_{M \in \cp} M$ to $[0,1]$ such that for any $M$,
$\sum_{x \in M} \omega(x) =1$.  
\end{definition}
$M$ are the possible measurements;
taking them to be disjoint means we are not allowing any {\em a
priori} identification of outcomes of different measurement
procedures.  $\Omega$ is the set of phenomenologically
admissible compendia of probabilities for measurement outcomes.  
The set $\cm$ is an
example of what Foulis calls a ``test space'': a set $\ct$ of sets
$T$, where $T$ may be interpreted as operations, (tests, procedures,
whatever you want to call them) and the elements $t \in T$ as outcomes
of these operations.  
(Without the interpretation, these are better known
in mathematics as {\em hypergraphs} or {\em set systems}.)  
Call the set of all outcomes $\Lambda := \union T$.
In general test spaces
the $T$ need not be disjoint; here they are. Foulis calls such test
spaces ``semiclassical.''
(Sometimes a weak requirement of {\em
irredundancy}, that none of these sets is a proper subset of
another, is imposed on test spaces; it is automatic here.)
States on test spaces are
functions $\omega: \Lambda \rightarrow [0,1]$
such that $\sum_{t \in T} \omega(t) = 1$ for any $T$.  
It is only when a phenomenological theory is defined
as a set of states on a general $\ct$, where a given outcome may occur in
different measurements, that the question of contextuality (does the
probability of an outcome depend on the measurement it occurs
in?) arises at the phenomenological level.  By not admitting such a
primitive notion of ``same outcome,'' but distinguishing outcomes
according to the measurements they occur in, the construction we 
make will 
guarantee noncontextuality of probabilities even at
the later stage where the theory is represented by a more abstract
structure in which the elements (effects, or operations) that play the
role of outcomes may occur in different operations.  
Though the
rest of our discussion ignores it, the question of whether there
can be convincing reasons for admitting a primitive notion of
``same outcome'' (based perhaps on some existing theory in terms of
which the operations and experiments of our ``phenomenological
theory'' are described) is worth further thought.  A related point is
that test spaces provide a framework in which we can implement a
primitive notion of two outcomes of different measurements being the
same, but we cannot implement a notion of two outcomes of {\em the
same} measurement being the same (up to, say, arbitrary labeling).  
A formalism in
which one can is that of E-test spaces (the E is for effect).
These are sets, not of sets of outcomes, but of multisets of outcomes.
Multisets are just sets with multiplicity: each element of the
universe is not just in or out of the set, but in the set with a
certain nonnegative integer multiplicity.  Where sets
can be described by functions from the universe to $\{0,1\}$ (their
characteristic functions), multisets are described by functions from
$U$ to $\N$.  The set of resolutions of unity in an effect
algebra, shorn of its algebraic structure, is an E-test space
(whence the name).  Not all E-test spaces
are such that an
effect algebra can be defined on them; those that are are called {\em
algebraic}.  Sufficiently nice E-test spaces are {\em prealgebraic},
and can be completed to be algebraic by adding more multisets
without enlarging the universe (underlying set of outcomes).

To each phenomenological theory we may associate a set of 
Boolean algebras, one for each measurement.  We will call this
set of Boolean algebras 
the ``phenomenological logic'' of the theory;  note,
though, that it is independent of the state-set  $\Omega$. 
These are just the
subset lattices of the sets $M$, or what I previously called the
``propositional logics'' of statements about the results of the measurements. 
We will distinguish them by subscripts on the connectives saying which 
measurement is referred to, e.g. ${\wedge}_M$ (although this is redundant
due to the disjointness of the measurements).

The phenomenological states $\omega$ of $\cp$ 
naturally induce states (which we will also call $\omega$) 
on the logic of $\cp$, 
via $\omega(\{a\}) = \omega(a)$,
$\omega( X) = \sum_{x \in X} \omega(x)$.  They will 
satisfy $\omega(M) = 1$ for each $M$, and $\omega(\emptyset) = 0$.  
We have, for
example ($x$ and $y$ are now subsets of outcomes),
$
\omega(x 
{{\vee}_M} 
y) = \omega(x) + \omega(y) - \omega(x 
{\wedge}_M 
y),
$
(which is equivalent to its dual).
We call the elements of the Boolean algebras of a phenomenological logic 
{\em events}, and we will refer to the set of events of $\cp$
as $\cv$.

\begin{definition}
Events $e, f$ are {\em probabilistically
equivalent}, $e \sim f$  in a phenomenological theory if they have the 
same probability under all states:
$
\forall \omega \in \Omega, \omega(e) = \omega(f)\;.
$
\end{definition}

$\sim$ is obviously an equivalence relation (symmetric, transitive,
and reflexive).  Hence we can divide it out of the set $\cv$,
obtaining a set $\cv/\sim ~=: \ce(\cp)$ of equivalence classes of
events which we will call the {\em effects} of the theory $\cp$.  (We
have dependence on $\cp$, rather than just $\cm$, because although
$\cv$ depends on $\cm$ but not $\Omega$, $\sim$ depends also on
$\Omega$. ) Call the canonical map that takes each element $a \in \cv$
to its equivalence class, ``$e$.''
The images $e(M)$ of the measurements $M$ under $e$ are ``measurements 
of effects.''  Together they form an $E$-test space as defined above
(a set of multisets). 
We now define on this space 
another ``logic'' which is, at least as far as possible, the 
simultaneous 
``image'' under the map $e$ of each of the Boolean algebras $M$.
To this end,  we introduce a binary operation $\oplus$ on the 
effect space.
\begin{definition}
\label{jammin}
$e_1 \oplus e_2 := e(a \vee_M b)$
for some $a$ such that $e_1= e(a)$, $b$ such that $e_2 = e(b)$, 
and $M$ such that $a,b \in M$ but $a \cap b = \emptyset$.
\end{definition}
If no such $a,b,M$ exist, 
$\oplus$ is 
undefined on the effect space.
(If they do exist, we will say they {\em witness} the existence
of $e_1 \oplus e_2$.)
As part of the proof of Theorem \ref{theorem: main}
we will show from the definition of the map $e$ via 
probabilistic equivalence and the behavior of probabilities with 
respect to $\vee_M,$ that this definition is independent of 
the choice of $a,b,M$.

Let $\omega^e$ denote
the function from the effects to $[0,1]$ induced in the obvious
way by a state $\omega$ on the Boolean algebra:  effects being
equivalence classes of things having the same value of $\omega$,
we let $\omega^e$ take each equivalence class to $\omega$'s value
on anything in it.  

\begin{definition}
A set of states $\Omega$ on a WEA $\ce$ is {\em separating} if
for $x,y \in \ce, ~
x \ne y \Rightarrow \exists \omega \in \Omega 
(\omega(x) \ne \omega(y))$.
\end{definition}

\begin{theorem} \label{theorem: main}
The set $\ce(\cp)$ of effects of a phenomenological theory $\cp$
with state-set $\Omega$, 
equipped with 
the operation $\oplus$ of Def. \ref{jammin} and the definition
$1 = e(1_M)$ (for some $M$) constitutes a weak 
effect algebra.  There exist phenomenological theories for which
this is properly weak, i.e. not an effect algebra.  
For all $\omega \in \Omega$
the functions 
$\omega^e$ defined above are states on the resulting weak effect
algebra.  $\Omega^e := \{\omega^e | \omega \in \Omega\}$ is separating on 
$\ce(\cp)$.
\end{theorem}

The proof is a straightforward verification of the axioms and 
the statements about states from 
the definition, and an example for the second sentence.

\begin{proof} \noindent
We begin by demonstrating 
$\oplus$ is in fact a partial binary
operation.  This is done by verifying the 
independence, asserted above, of the definition
of $\oplus$ from the choice of $a,b,M$ 
and of $1$ from $M$.  Suppose  $e_1 = e(a)= e(c),
e_2 = e(b) = e(d), a,b \in M, c,d \in N, a \ne b,
c \ne d,$ $a \wedge_M b = 0$, $c \wedge_N d = 0$.
Consider any state $\omega$ on 
the set of Boolean algebras which is also in $\Omega$, the
states of our phenomenological theory.  By 
the definition of $e$, 
$\omega(a) = \omega(c) {\rm~~and~~}\omega(b) = \omega(d)\; ;
$
therefore $\omega(a) + \omega(b) = \omega(c) + \omega(d)$.
Now $\omega(a \vee_M b) = \omega(a) + \omega(b)$ because
$a \vee_M b = 0$, and similarly $\omega(c \vee_N d) 
= \omega(c) + \omega(d)$. 
In other words, for any state $\omega \in \Omega$, 
$\omega(a \vee_M b) = \omega(c \vee_N d)$, so
$a \vee_M b$ and $c \vee_N d$ are probabilistically equivalent, and
correspond to the same effect.

Each Boolean algebra contains a distinguished element
$1$;  by the definition of state on $\cp$,
these have probability zero, and one, respectively, in all states.
Hence they each map to a single effect, and these effects we will
call $0$ and $u$ in the effect algebra (verifying later that
$0 = 1'$ in the weak effect algebra, so that it is consistent with
the usual definition of $0$ in a WEA).   Of course,
$\omega^e(1) =1$.  It is also easy to see that 
$\omega^e(x \oplus y) = \omega^e(x) \oplus \omega^e(y)$.
Hence the $\omega^e$ are states, as claimed.  
The set $\Omega^e$ is obviously separating.  
To be pedantic, suppose 
there exist effects $x,y$ having $\omega^e(x) = \omega^e(y)$ for
all $\omega^e \in \Omega^e$.  By the definition of $\omega^e$, 
$\omega^e(x)$ is    the common value of
$\omega$ on all $e$-preimages of $x$, and $\omega^e(y)$ is the 
common value of $\omega$ on all $e$-preimages of $y$.  If these 
values are the same for all $\omega^e$, then the preimages of $x$
and of $y$ are all in the same equivalence class, so $x=y$.  
Hence, $\Omega^e$ is separating.

We now verify that $\oplus$ satisfies the weak 
effect algebra axioms.

(EA1) Strong commutativity:  If $a,b \in M$ witness the existence of 
$x \oplus y$ as described in the definition of $\oplus$, 
by symmetry of $\vee_M$ and $\wedge_M$ (which enter symmetrically
in the definition of $\oplus$) 
they also witness the existence of $y \oplus x$
and its equality with $x \oplus y$.  

(WEA2) Weak associativity.
Let $a,b \in M, e(a)=x, e(b) =y, a \cap b = \emptyset$, so that
$a,b$ witness the existence of
$x \oplus y$, and also let 
$c, d \in N \text{ and disjoint, } 
e(c) = z, e(d) = x \oplus y$, so $c,d$ witness the existence
of $(x \oplus y)\oplus z$.  
Similarly let $b',c' \in P$ 
witness the existence of 
$y \oplus z$ and 
$a',f \in Q$  witness
the existence of $x \oplus (y \oplus z)$, so that 
$e(a') = x, e(f) = y \oplus z$, and $a',f$ are disjoint.
Then $\omega^e(x \oplus y) = \omega(a) 
\oplus \omega(b)$ and 
$
\omega^e( (x \oplus y)) \oplus z)
= \omega(a) + \omega(b) + \omega(c)\;.
$
Also $\omega^e(y \oplus z) = \omega(b') \oplus \omega(c') = 
\omega(b) \oplus \omega(c)$, 
so 
$
\omega^e((x \oplus (y \oplus z)) = \omega(a') \oplus \omega(f)
= \omega(a) \oplus \omega(b) \oplus \omega(c)\;.
$
But $\omega^e((x \oplus y) \oplus z) = \omega^e(x \oplus (y \oplus z))$
for all $\omega^e$ implies $(x \oplus y) \oplus z = 
x \oplus (y \oplus z)$ by the fact that $\Omega^e$ is separating.

(EA3) Define $e'$ to be $e(a')$, for any $a$ such that $e(a) = e$, 
and $a'$ is $a$'s unique complement in the Boolean algebra of 
the measurement $M$ containing it.   
Since for any state, 
$\omega(a')=1-\omega(a)$ and this probability is independent of $a$
as long as $e(a)=e$, $e'$ as thus defined is independent of which 
$a$ is chosen.  Moreover, since $a \wedge_M a' = 0$ 
$e \oplus e' \equiv e(a) \oplus e(a')$ is defined and
equal to $e(a \vee_M a') = e(1_M) = 1,$ so that $'$ as we 
just defined it satisfies (EA3).  

(EA4)  Note that $x \oplus 1$ is equal to $e(a \vee_M 1_M)$, 
for some $M$ containing $a$ and with unit $1_M$, where 
$a \wedge_M 1 = 0$ and  $e(a)=x$.
But each $M$ has a unique $a$ such that $a \wedge_M 1_M = 0_M$, 
namely $0_M$. So an $x$ such that $x \oplus 1$ exists;  it
must be $e(0_M)=0$. 

This proves the first part of the theorem.  
We remark that $1' \equiv e(1') = 
e(0_M)$, so defining $0$ as $e(0_M)$ for
any $M$ coincides with the usual effect algebra definition
as $1'$.  
We now construct 
the counterexample required by the second part.

Consider a phenomenological theory
consisting of states on the two atomic Boolean algebras:
\beqa \label{example}
\begin{array}{lcccccccccr}
M:& ( & a & & b & )&( & & f &  & ) \\
N:& ( &  & c & & )&(& d ~~~)& ( & g & )
\end{array}
\eeqa
with the indicated $a,...,g$ being atoms of the
Boolean algebras involved (``elementary measurement outcomes''). 
The vertical lining-up of 
parentheses in (\ref{example})
visually indicates conditions we will impose on the theory:
that all states of our phenomenological theory respect 
$\omega(a \vee_M b) = \omega(c)$ and $\omega(f) = \omega(d \vee_N g)$;
further, let our theory contain states with nonzero probability for each of
$a,b,c,d,f,g$.  
There are plenty of perfectly good empirical theories satisfying
these constraints, but $\oplus$ on the effect set of such a theory
will not exhibit strong associativity:
although $e(a) \oplus e(b)$ exists and is equal to $e(c)$, 
and $e(c) \oplus e(d)$ exists and is therefore equal to 
$(e(a) \oplus e(b)) \oplus e(d)$, no effect $h$ exists
with $e(h) = e(b) \oplus e(d)$.
\end{proof}

\begin{conjecture}[Completion conjecture for WEA's]
Let $\ce$ be a WEA obtained from a phenomenological theory.
A unique effect algebra $\overline{\ce},$ which we call the
{\rm completion} of $\ce$,  
can be constructed from $\ce$ as follows.  Whenever
only one side of the associativity equation exists, 
impose the equation (extend $\oplus$ to contain the pair that would 
appear on the other side).
This can also be characterized as the smallest effect algebra containing
$\ce$ as a sub-weak-effect-algebra (with the latter concept appropriately
defined).
\end{conjecture}
Thus the well-developed and attractive theory
of effect algebras could be useful in this more general context.
The adjunction of these new relations and the new resolutions of 
unity whose existence they imply is an interesting theoretical move.
In constructing theories, we often suppose the existence of things that
do not, at least initially, correspond to things in the available 
phenomenology.  The idea of including all Hermitian operators as observables
in quantum mechanics is an example; there has been
much discussion of whether they are all operationally
observable.  This has motivated the search, often successful, for methods
of measuring observables that had previously not been measured,
and the development of a general theory of algorithmic procedures for
measurement.  The conjecture above might motivate the search
for empirical methods of  making measurements which would correspond 
to the additional resolutions of unity needed
to make the initial WEA into an effect algebra.  In any case, it is
worth studying the nature of information processing  
and information theory (if the latter still makes sense) in 
properly weak effect algebras versus their 
completions.

We are now ready for a few remarks on the significance of Gleason's theorem 
\citep{Gleason57a} in
this context.
Gleason's
theorem says that 
in Hilbert space dimension greater than two,
if mutually exclusive
quantum measurement results are associated with mutually orthogonal
subspaces of a Hilbert space, 
and exhaustive sets of such measurements
to direct sum decompositions of the space into such subspaces,
and if the probability of getting the result associated to a given
subspace in a given measurement is independent of the measurement
in which it occurs 
(``noncontextual'') then the probabilities must be
given by the trace of the product of the projector onto the given
subspace with a density operator.
A similar theorem resolutions of unity into orthogonal
projectors replaced by resolutions into arbitrary positive operators
has been obtained by \cite{Busch99a}, and independently by Caves,
Fuchs, Mannes, and Renes \citep{Fuchs2001a, Fuchs2002a}.  In the next section
we will see how this theorem is a case of a general fact about convex
effect algebras.

Sometimes Gleason-type theorems are used to justify the quantum probability law. Then one 
must justify the assumptions that
probabilities are noncontextual, and that they are associated with
orthogonal decompositions, or positive resolutions of unity, on a
Hilbert space.  Although Theorem \ref{theorem: main} 
gives structures (WEAs) much more general than Hilbert space
effect algebras (or their subalgebras consisting of projectors),
it automatically results in noncontextual probability laws.  
But he construction of WEAs in Theorem 1 starts from
probabilities, so it would be circular to use it to justify
noncontextuality in an appeal to Gleason's theorem to establish
quantum probabilities.  Rather, Theorem \ref{theorem: main} says
that we can elegantly, conveniently
represent {\em any} empirical theory by a set of noncontextual
probability assignments on a certain WEA (and, if the
completion conjecture is correct, embed this in an effect algebra).
In the case of quantum theory, this general recipe provides {\em both}
the Hilbert space structure {\em and} the trace rule for
probabilities, as a {\em representation} of the compendium of
``empirical'' probabilities (perhaps somewhat idealized by the
assumption that any resolution of unity can be measured) of quantum
theory.

The generalization of Gleason-like theorems to weak effect algebras,
effect algebras, and similar structures are theorems characterizing
the full set of possible states on a given such structure, or class of
such structures.  
In the particular case of a Hilbert
space effect algebra, the import of the B/CFMR theorem, from our
operational point of view, is that the quantum states
constitute the {\em full} state space of the ``empirically derived''
effect algebra.  This is especially interesting since
in other respects, the category of effect algebras
probably does not have enough structure to capture everything we would
like it to about quantum mechanics: for example, the natural
category-theoretic notion of tensor product of effect algebras
(\cite{Dvurecenskij95a}; see also \cite{Wilce94a,Wilce98a}), applied
to effect algebras of finite dimensional Hilbert spaces, does not give
the effect algebra of the tensor product Hilbert space (or of any
Hilbert space), as one sees from a result in \cite{Fuchs2001a} (a
similar result involving projectors only is in \cite{Foulis81a}).
Possibly relatedly, a natural category of morphisms for convex effect
algebras, those induced by positive (order-preserving) linear maps on
the underlying ordered linear space (see below), is larger in the
quantum case than the ``completely positive'' maps usually considered
reasonable for quantum dynamics.  Nevertheless for a given Hilbert
space effect algebra, its set of all possible
states is precisely the set of quantum states.

The role of Gleason-like results depends to some extent on point of
view.  In the project of exploiting the analogies between quantum
states and Bayesian probabilities, they can play a nice conceptual role.  
Probabilities are, roughly, ``the right way''
(nonarchimedeanity issues aside) to represent uncertainty, and to
represent rational preferences over uncertain classical alternatives.
In this ``Bayesian'' project, it would be very desirable to see quantum states as
``the right way'' to deal with uncertainty in a nonclassical
situation: the Hilbert space structure perhaps sums up the
``nonclassicality of the situation,'' and the probabilities can be
seen as just the consequence of ``rationality'' in that situation.
This suggests that the ``structure of the nonclassical situation'' mentioned
above might be described in terms of measurement outcomes (sometimes
called ``propositions'' or ``properties'') having probability zero or one;  
then Gleason's
theorem or analogues for other ``property'' structures, might give the
set of possible probability assignments for such a structure.  
This is related to the ``Geneva'' approach to empirical theories 
(rooted in the work of Jauch and Piron on ``property lattices'').

\subsection{Convex effect algebras}
It is natural to take the
space of operations one may perform as convex.  This represents
the idea that given any operations $M_1$ and $M_2$,
we can perform the operation $(\lambda_1 M_1, \lambda_2 M_2)$ (where
$\lambda_i \ge 0, \lambda_1 + \lambda_2 = 1$) in which we perform one
of $M_1$ or $M_2$, conditional on the outcome of flipping a suitably
weighted coin (or, in more Bayesian terms, arrange to believe
that these will be
performed conditional on mutually exclusive events, to which we assign
probabilities $\lambda_1, \lambda_2$, that we believe to be
independent of the results of measurements on the system under investigation).  
If we looked at the coin face and saw the
index ``$i$'' and obtained the outcome $a$ of $M_i$, this should
correspond to an outcome $\lambda a$ of $(\lambda_1 M_1, \lambda_2
M_2)$, and any state should satisfy $\omega(\lambda a) = \lambda
\omega(a)$.

Similar assumptions may be made at the level of effect algebra.  
For effect algebras
constructed via probabilistic equivalence,
they will be consequences of 
the convexity assumptions on the initial phenomenological theory; 
this will be worked out elsewhere.  
One could also
pursue the consequences of imposing a generalized convexity based on a
more refined notion of ``vector probabilities'', or other
representations of uncertainty by nonarchimedean order structures.
Such generalized probabilities and
utilities
can result from Savage-like representation theorems for preferences
satisfying ``rationality'' axioms but not certain technical axioms
that make possible real-valued
representations \citep{LaValle92a,LaValle96a,Fishburn98a}.  
We will avoid such complications, but
knowing about them may clarify the role of some technical
conditions in results to be discussed below.

\begin{definition}
A {\em convex effect algebra} is an effect algebra $\langle E, u,
\oplus \rangle$ with the additional assumptions that for every $a \in
E$ and $\alpha \in [0,1] \subset \R$ there exists an element of E,
call it $\alpha a$, such that  (C1) $\alpha(\beta a) = (\alpha
\beta) a$,  (C2) If $\alpha + \beta \le 1$ then $\alpha a \oplus
\beta a$ exists and is equal to $(\alpha + \beta) a$,  (C3) $\alpha
(a \oplus b) = \alpha a \oplus \alpha b$ (again, the latter exists),
 (C4) $1a=a$.
The mapping $a \mapsto \alpha a$ from $[0,1] \times E$ to $E$ is
called the {\em convex structure} of the convex effect algebra.
\end{definition}

\citet{Gudder98b} 
showed that ``any convex effect algebra
admits a representation as an initial interval of an ordered
linear space,'' and in addition if the set of states on the
algebra is separating, the interval is generating.  
To understand this result, we review the mathematical
notion of a ``regular'' positive cone (which we will just call cone);
it is basic in quantum information science, e.g because the
quantum states, the
separable states of a multipartite quantum system, the
completely positive maps, the positive maps, unnormalized in each case,
form such cones.

\begin{definition}
A {\em positive cone} is
a subset $K$ of a real vector space $V$ closed under multiplication by
positive scalars.
It is called {\em regular} if it is (a) convex (equivalently,
closed under addition:  $K + K = K$), 
(b)
 generating ($K-K=V$, equivalently $K$
linearly generates $V$,) 
(c) 
pointed ($K \cap -K = \emptyset$, so that it
contains no nonnull subspace of $V$), and (d) 
topologically closed (in the
Euclidean metric topology, for finite dimension).  
\end{definition}


Such a positive cone induces a {\em partial order}
$\ge$ on $V$, defined by $x \ge_K y := x - y \in K$. 
$(V,\ge_K)$,
or sometimes $(V,K)$,  is
called an {\em ordered linear space}.  The Hermitian operators on a
finite-dimensional complex vector space, with the 
ordering induced by the cone of positive semidefinite operators, are
an example.  (A relation $R$ is defined to be a partial order
if it is reflexive ($x R x$), 
transitive ($x R y ~\&~ y R z \Rightarrow x R z$) and antisymmetric
($(x R y ~\&~ y R x) \Rightarrow x=y$.)  The partial orders induced
by cones have the property that they are ``affine-compatible'':
inequalities can be added, and multiplied by positive scalars.  If
one removes the requirement that the cones be generating, cones
are in one-to-one correspondence with affine-compatible partial orderings.
In fact, the categories of real vector spaces with distinguished
cones, and partially ordered linear spaces, are equivalent.

We pause to motivate
some of the seemingly technical
conditions of regularity.  A regular cone may represent the
set of unnormalized probability states of a system, or a set of
specifications of expectation values of observables.  
The normalized states may be
generated by intersecting it with an affine plane not containing the
origin.  Convexity is fairly clearly motivated by operational
considerations, such as those in the definition of convex effect
algebra above, or in the desire to have a normalized state set given
by intersecting the cone with an affine hyperplane be convex.  Topological
closure is required so that the cone has extreme rays, and the convex
sets we derive by, for instance, intersecting it with an affine hyperplane,
will have extreme points if that intersection is compact;  then the
Krein-Milman theorem states that these extreme points convexly generate
the set.  (An affine hyperplane is
just a translation of a subspace: for $d=3$, a $2$-d hyperplane is a plane
in the sense of high school geometry.)  In ``empirically motivated''
settings such as ours, in which the metric on the vector space will
be related, via probabilities, to distinguishability of states or
operations, limit points can be as
indistinguishable as you want from things already in the cone, so
closing a cone cannot have empirically observable effects, and may as
well be done if it is mathematically convenient.  In the presence of
some of the other assumptions, pointedness ensures that the
intersection with an affine plane can be compact.  Its appearance in
the representation theorem for convex effect algebras (presumably
essentially because the convex sets one gets via states tend to be
compact intersections of an affine ``normalization'' 
plane with such a cone) is one
``operational'' justification for pointedness. Pointedness also has a
clear geometric interpretation: if the subspace $K \cap -K$ is one-dimensional,
instead of a ``point'' at zero the cone could
have an ``edge,'' which is why nonpointed
cones are often referred to as ``wedges''; of course $dim(K
\cap -K) > 1$ is also possible for a nonpointed cone.  
The property of being generating is often
appropriate because any non-generating cone generates a
subspace, and we may as well work there.
When several cones are considered at once, this might no longer be
appropriate.

An {\em initial interval} in such a space is an interval 
$[0, u]$ defined as the set of things between zero and $u$ in 
the partial ordering $\ge_K$, i.e. \
$\{ x \in V: 0 \le_K x \le_K u\}$. It is generating if it
linearly generates $V$. 
It can be viewed as a convex effect algebra by letting $\oplus$ be 
vector addition restricted to $[0,u]$ and the convex structure
be the restriction of scalar multiplication.
The representation theorem says any 
convex
effect algebra is isomorphic (as a convex effect algebra) to some
such linear convex effect algebra (via an affine map).
In 
finite-dimensional 
quantum mechanics the vector space and cone are $H_{d}$ and
the positive semidefinite cone, and
the interval referred to in the representation
theorem is $[0, I]$.

In addition to the requirements for states on an effect algebra,
states on a convex effect algebra must satisfy $\omega(\lambda a)
= \lambda \omega(a)$.  
The set of all possible states on a convex effect algebra may 
be characterized via a version of 
Lemma 3.3 of 
\cite*{Gudder99b}, which describes it for linear
effect algebras $[0,u]$.  First, some definitions.
The dual vector space $V^*$ for real $V$
is the space of linear functions (``functionals'') 
from $V$ to $\R$;  the dual cone $K^*$ (it is a cone in $V^*$) 
is the set of linear functionals which are nonnegative on $K$.
Then $\Omega([0,u])$, the set of all states on $[0,u]$ when 
the latter is viewed as a convex effect algebra, is precisely
the restriction to $[0,u]$ of the set of linear functionals 
$f$ positive on $K$ and with $f(u)=1$ (``normalized'' linear
functionals).    The restriction map is a bijection.
Viewing things geometrically,  the states (restricted functionals)
are in one-to-one correspondence with
the (unrestricted) 
functionals in the intersection of $K^*$ with the
affine plane in $V^*$ given by $f(u)=1$.
Since 
any linear functional on the $d^2$-dimensional vector
space $H_{d}$ of Hermitian operators on $\C^d$
has the form 
$X \mapsto \tr A X$ for some $A$,
while the dual to the positive semidefinite cone in 
$H_{d}$ is the set of
such functionals for which $A \ge 0$
(i.e., the positive semidefinite cone is self-dual ($K=K^*$))
this Lemma tells us that 
the states of 
a finite-dimensional Hilbert space effect algebra are precisely those
obtainable by tracing with density matrices $\rho$; in other words, 
the Gleason-type theorem for POVMs is a case of this general
characterization of states on convex effect algebras.  This
illustrates
the power and appropriateness of this approach (and probably other convex 
approaches, in which similar characterizations probably exist) 
to empirical theories, and to problems in quantum 
foundations.  Gleason's theorem itself cannot be established in this
way, because the effect algebra  of 
projectors is not convex.  However, there may be a natural 
notion of ``convexification'' of effect algebras 
according to which $[0,I]$ is the convexification
of the effect algebra of projectors.  Interesting questions are then,
which effect algebras can be convexified, and for 
which of those (as for the effect algebra of quantum projectors)
convexification does not shrink the state-space.  
Conversely, we might ask for ways of identifying special subalgebras 
of effect algebras, composed of effects having special properties 
like ``sharpness'', perhaps having additional structure such as that
of an orthoalgebra, and investigate the relation between state-sets of 
effect algebras and these sub-algebras.

\subsection{Sequential operations}

The operational approach I am advocating suggests that one consider
what general kinds of ``resources'' are available for
performing operations.  Provided both system and observer are
sufficiently ``small'' portions of the universe, it may be reasonable
to suppose that the observer may use yet other subsystems (distinct
from both observer and system) as an ``apparatus'' or ``ancilla'' to
aid in the performance of these operations, that the apparatus may be
initially independent of the system and observer, and that the
combination of apparatus and system may be viewed as a system of the
same general kind as the original system, subject to the same sort of
empirical operational theory, with a structure, and a state, subject
to certain consistency conditions with that of the original system.
(Convexity is a case of this, the ancilla functioning as ``dice.'')
It may be that in some limits some of these assumptions break down,
but it is still worth investigating their consequences for several
reasons: so that we can recognize breakdowns more easily, so that we
may even acquire a theoretical understanding of when and why to expect
such breakdowns, and because we may gain a better understanding 
of why empirical theories valid in certain limits (say, 
small observer, 
small apparatus, small system) have the kind of structure they do.

Besides convex combination, 
other such elementary combinations and conditionings of operations
should probably be allowed:  essentially, the set of operations should
be extended to allow including them as subroutines
in a classical randomized computation.  (Of course, this will not
always be appropriate;  for example, in investigating or constructing
theories that are not even classically computationally universal.)
Among other things, this
might get us the $\oplus$ operation previously obtained as the image
of OR($\vee$) in Boolean propositional logics about each operation's outcomes,
``for free,'' as we can use classical circuitry to construct procedures
whose outcomes naturally correspond to propositional combinations
of the outcomes of other procedures, 
and will have the same probabilities as those combinations.
This leads us to the consider the possibility that the set of 
possible operations be closed under {\em conditional composition}.
This means that given any operation $M$, and set of operations
$M_\alpha$, $\alpha \in M$, there is an operation consisting of
performing $M$, and, conditional on getting outcome $\alpha$ of
$M$, then proceeding to perform $M_\alpha$.  
This assumption is
natural, but nevertheless substantive:
one could imagine physical theories that did not satisfy it.
Some outcomes might destroy the system, or so alter
it that we can no longer perform on it all the procedures we could
before.  Nevertheless, it is worth investigating the structure
of theories satisying the assumption (the theory of quantum operations being
one such case).  The 
structures obtained when conditional composition is not universally
possible might turn out to be understandable as partial versions
of those we obtain when it is always possible, or in some other
way be easier to understand once the case of total conditional 
composability is understood.
An operation in this framework, then, can be viewed as a
tree with a single root node on top, 
each node of which is labelled by an operation and the 
branches below it labelled by the outcomes of the operation,
except that the leaves are unlabelled (or redundantly labelled
by the labels of the branches above them).  The interpretation
is that the root node is the first operation performed, and
the labels of the daughters of a node indicate the operation 
to be performed conditional on having just obtained the 
outcome which labels the branch leading to that daughter. 

From now on, we mean by phenomenological theory a {\em sequential} 
phenomenological theory, i.e. one closed under conditional composition.
If we extend a
phenomenological theory via this requirement, the new
outcome-set contains all finite strings of elements of the old outcome
set.  Given closure under conditional composition, a given string can
now appear in more than one measurement.  In order that the
construction of dividing out operational probabilistic
equivalence can work,
we will have to require that the empirical probability of the string
be noncontextual.  We will also use a different notion of
probabilistic equivalence: $x \sim y$ iff for any $a,b$, $\omega(a x
b) = \omega(a y b)$, where $x,y,a,b$ are outcome-strings.  In our context
the noncontextuality assumption can actually be derived from
the disjointness of ``elementary''
operations (those not constructed via composition) and the assumption
that the choice of operation at node $n$ 
of the tree describing an operation constructed via conditional 
composition cannot affect the probabilities of outcomes corresponding
to paths through the tree not containing node $n$.  This is how one 
might formalize a  generalization of the ``no Everett phone'' requirement
suggested in Polchinski's 
\cite*{Polchinski91a}
article on Weinberg's nonlinear
quantum mechanics:  the probability of an outcome
sequence cannot depend on what operation we {\em would have done} had
some outcome in this sequence not occurred.   

With suitable additional formalization of the notion of phenomenological
operational theory, and appropriate definitions of $\oplus$ and a 
sequential product on the resulting equivalence classes, one can prove
that dividing probabilistic equivalence out of such a
set of empirical operations, in a manner similar to the
construction of weak effect algebras via probabilistic equivalence,
gives what I will call a {\em weak operation algebra}.  The details
will be presented elsewhere.  Here
I will exhibit the quantum-mechanics of operations as a case of 
a general structure, an {\em operation algebra} (OA), 
which I view as the analogue, for operations,
of an effect algebra.  The structure will be related to the
notion of sequential effect algebra (SEA) studied by 
\citet*{Gudder2000a}, but differ from it in important
respects.  It would be interesting to study when the set of 
effects of an OA forms a SEA.  

Since this structure will be a partial abelian semigroup, with extra
structure involving only the PAS operation $\oplus$, with a product
meant to represent composition of operations, and additional axioms
about how the two interact, we will discuss some more aspects of PASes
(following \cite{Wilce98a}) before defining operation algebras.  The
reader might want to keep in mind the algebra of trace-nonincreasing
completely positive maps (with $\oplus$ as addition of maps and the
product as composition of maps) as an example.

Recall that a PAS is a set with a strongly commutative and strongly
associative partial binary operation $\oplus$ defined on it.  Define a
zero of a PAS as an element $0$ such that for any $a$, $a \oplus 0 =
a$. (Uniqueness follows.)  If a PAS does not have a zero, it is
trivial to adjoin one; we henceforth include its existence as part of
a PAS.  A PAS is {\em cancellative} if $x \oplus y = x \oplus z
\Rightarrow y = z$, {\em positive} if $a \oplus b = 0 \Rightarrow
a,b=0$.  The relation $\le$ on a PAS is defined by $x \le y
\Leftrightarrow \exists z~~ x \oplus z = y$.  Part of Lemma 1.2 of
\cite{Wilce98a} is that in a cancellative, positive PAS $\le$ is a
partial ordering.  In such a PAS, we define $T$ as the set of top
elements of the partial ordering (i.e. $T = \{ t \in \co | a \oplus t
\text{ exists } \Rightarrow a = 0\}$).  In a cancellative PAS we
define $x \ominus y$ as that unique (by cancellativity) $z$, if it
exists, such that $y \oplus z = x$.  Define a {\em chain} in 
a partially ordered set $P$  as a set $C \subseteq P$
such that $\le$ restricted to $C$ is total.

\begin{definition}
An {\em operation algebra} $\co$  is a 
cancellative, positive PAS  equipped with a total
binary operation, the sequential 
product, which we write multiplicatively.  With respect
to the product, the structure is
(OA5) a monoid (the product is associative) 
with (OA6) a unit $1$ (semigroup is sometimes used as 
a synonym for this 
unital monoid structure). The remaining axioms
involve the interaction of this monoid structure with the PAS structure.
\\
(OA7) $0c=c0=0$.\\
(OA8) $(a \oplus b)c = ab \oplus bc$, 
$a(b \oplus c) = ab \oplus ac$ (distributive laws).\\
(OA9) $1 \in T$.\\
(OA10) Every chain in $\co$ has a sup in $\co$.
\end{definition}

Note that the sup mentioned in (OA10) is not necessarily in the chain.
(OA10) says that $\co$ is {\em chain-complete}; this is (nontrivally,
and I am not certain whether choice or other strong axioms are
required in the infinite case) equivalent to saying it is {\em
complete}, meaning that every directed subset of $\co$ has a $\sup$ in
$\co$.  (A poset $P$ is directed if for every subset $S$ of it, $P$
contains an element $x$ greater than or equal to everything in $S$.)
The thinking behind
(OA10) is that we are to conceive of the elements or ``operations'' in
$\co$ as possible outcomes of procedures performed on a system, and
each such outcome must be part of at least one exhaustive set of 
such outcomes.  Given how the ordering is defined, it might seem 
natural therefore to require that all upward chains terminate; 
however, when there are sufficiently many operations (and also, 
but not only, if continuous sets of outcomes for a given operation
are envisaged), as in the quantum case, it could be reasonable
to allow (what is certainly possible in the quantum case) chains
that do not terminate, but have a limit point (the sup mentioned
in (OA10)).

Our structure is not an effect algebra because we do not assume it is
(as a PAS) unital (i.e., has at least one unit).  
A {\em unit} of a PAS is an element $u$ such that
for any $a$, there is at least one $b$ such that $a \oplus b = u$.  In
a cancellative, positive, unital PAS (equivalently, effect algebra)
there is a unique unit (the sole element of the top-set $T$).
Axiom (OA10) might need strengthening in order to obtain some of 
the results one would like.  Notably, we would like to have a 
representation theorem in which the operations belong to a cone in 
a vector space (and thus belong to an {\em algebra} in one of the usual
mathematical senses, of a vector space with an appropriate product).
Aside from belonging to a cone, the special nature of the convex
set of operations in such a representation theorem would be expressed
by an additional requirement, deriving from (OA10),
which would specialize to the
trace-nonincreasing requirement in the case of the quantum operation
algebra (and generalize the initial interval requirement in the analogous
representation theorem for effect algebras).   

We shall now show that quantum mechanics provides an example of this
structure.  
We refer to the set of linear operators on $\C^d$ as
$B(\C^d)$.
\begin{proposition}\label{prop: qmeffect algebra}
The set of trace-nonincreasing completely positive linear maps on
$B(\C^d)$, with the identity map $\ci$
as $1$, the map $M$ defined by $M(X)=0$ for every $X$ as $0$, 
ordinary addition of maps as linear operators, restricted to the
trace-non-increasing interval, as $\oplus$, and composition of maps as
the sequential product, forms an operation algebra.  
Its top-set 
$T$ is the set of trace-preserving maps.
\end{proposition}

\begin{proof}
The commutativity (OA2) and associativity (OA1) of $\oplus$ and the
behavior of $0$ (OA7), and the unital monoid structure (OA5 and 6) are
immediate.  
Cancellativity holds for addition in 
any linear space, so since $\oplus$ is here a
restriction of addition on a linear space of linear maps, it is cancellative
(OA3).  It is positive (OA4) because $A + B = 0 \Rightarrow A,B = 0$ 
for $A,B$ in a pointed cone (such as the cone of completely positive linear
maps).   
(OA8) follows from the distributivity of multiplication of
linear operators over addition of linear operators.  The top-set $T$
is the set of trace-preserving operations, which follows from the
easy observation that if you add any operation besides the zero
operation to a trace-preserving operation, the result is not
trace-nonincreasing.  (OA9) follows since the identity operation is trace
preserving.  (OA10) involves an elementary topological argument which 
will be omitted here.
\end{proof}

We note the interpretation of $\oplus$ and $\ominus$ in terms of
the HK representation of a map $\ca$ in terms of operators $A_i$
(operators such that
the map acts as $X \mapsto \sum_i A_i X A_i^\dagger$).  Modulo
irrelevant details of indexing, the HK representation sequence $A_i$
is a multiset $[A]$ of operators $A$ such that $A^\dagger A \le 1$.
$\ca \ominus \cb$ exists if there are HK representations $[A], [B]$
such that $[B]$ is a submultiset of $[A]$.  (Equivalently, there
are standard HK representation sequences $A_i$ and $B_i$ such that
$B_i$ is an initial segment of $A_i$, i.e. $\cb(X) = \sum_i A_i X
A_i^\dagger$ where $i$ ranges over the first $k$ $A_i$.)  Thus it is
obvious that $\ca \ominus \cb$ will not always exist.

We define a {\em weak operation algebra} to satisfy all the above
axioms except that associativity is replaced with weak associativity
(whose statement is the same as in the definition of weak effect 
algebra).  With suitable additional formalization of the notion 
of sequential phenomenological theory and sequential probabilistic
equivalence, and definitions of $\oplus$
and sequential product on the equivalence classes, one can show:
\begin{theorem}
The set of equivalence classes obtained by dividing the 
notion of operational probabilistic equivalence defined above
out of a phenomenological operational theory, has a natural 
weak operation algebra structure.
\end{theorem}

Note that if we have operational limits on conditional composition, as
discussed above, we might accomodate that by modifying the notion of
operation algebra (or WOA) to make the multiplicative monoid structure
partial.  It would then be interesting to investigate the conditions
under which this partial structure is extendible to a total one (as
well as the conditions under which a WOA can be completed to an OA).

We can add a convex structure to an OA with little difficulty.
We just introduce a map of multiplication by scalars in $[0,1]$
(i.e. a map from $[0,1] \times \co \rightarrow \co$) such that
the axioms (C1--C4) of convex effect algebras hold, and also
$(\alpha a) b = \alpha (ab) = a (\alpha b)$ (COA15).
We expect such a structure to again emerge from an operational
equivalence argument applied to a suitable notion of convex operational
phenomenological theory.  

\section{Dynamics and the combination of subsystems in operational theories}

The operation algebra approach sketched above implicitly includes a
kind of dynamics, although without explicit introduction of a real
parameter for time.  Probably some operation algebras are extendible
to have a
notion of time.
However, in the quantum operation algebra given above the
assumption is that 
any completely
positive evolution can be achieved.  The time taken 
is neglected, and the temporal element of the interpretation
is only the primitive one that
when one measurement is done conditional on the result of another, it
is thought of as done after the result of the first is
obtained.
A more substantial notion of time might
be introduced in many different ways by adding structure to the
operation algebra, e.g. by some consistent specification of how long
each evolution takes, or by the assumption that each evolution can be
done in any desired finite amount of time.  The latter is a very
strong assumption.  In some cases, one might 
have a continuous semigroup structure related (with scheduling 
constraints) to
their sequential product.  A
 realistic consideration of these matters would
involve a much more detailed account of the interactions between
apparatus and system that are actually available.  This is an
important part of the project I propose, but I will not pursue it much
here.  It reminds us, though, of one of the important lessons of QIP for
foundations mentioned in Section \ref{sec: peculiar}: that which
operations are possible may depend on the resources available, and
that the beautiful structures one sometimes encounters as operational
theories may be idealized.  In particular, much of the attempt to
implement QIP involves struggling with the limitations imposed by the
limited nature of the subsystems, and interactions, physics makes
available.  It is important to incorporate such
limitations in operational structures.  
\cite{Barnum2002a} is one approach to this, with
the resources available for control and observation
limited (for example) to those definable 
via a Lie subalgebra of the full Lie algebra $\fsl(d)$
appropriate to arbitary quantum operations.  Physics 
includes much more than just Hilbert space: preferred bases or tensor product
structures, symmetries, the whole business of representation theory.
Another approach to involving this ``more'' 
in operational theories has been the
inauguration, particularly in works such as \cite{Foulis2000a} and
\cite{Wilce2000a}, of a theory of group actions on empirical
quantum logics.

An important part of the project of combining operational empirical
logic and QIP ideas to investigate whether or not physics can provide
an overarching structure unifying perspectives 
is to understand the operations available in an
operational theory in terms of interactions with apparatus and/or
environment.  In particular, if we have a way, such as the tensor
product in quantum mechanics, of describing the combination of
apparatus $A$ and system $S$ as subsystems of a larger system $L$, we
will probably want to require that the evolution induced on $S$ by
doing an operation on the larger system is, under appropriate
circumstances, one of the operations our theory describes as
performable on the smaller system.  ``Appropriate circumstances''
probably means that the apparatus should be initially independent of
the system, which in turn requires that the notion of combination of
subsystems have a way of implementing that requirement.  Such
assumptions bear close scrutiny, though, as they may be just the sort
of thing that becomes impossible in certain limits. Some, such as 
\citep{FOL2001a}, 
have argued for the physical relevance of
some situations in which open systems are analyzed without the initial
independence assumption.  Independence works well in the case of
completely positive quantum operations, though: indeed, all such
operations can be implemented via a {\em reversible} interaction with
apparatus.  
Consideration of categories, such as convex operation algebras
and generalizations of these, that describe dynamics is probably the
most promising way to investigate such questions.  
Possibly the category-theoretic notion of 
tensor product will be defined for these categories.  One could then
examine, for example, whether the tensor
product of two Hilbert-space CP-operation algebras is the operation algebra
of CP-maps on the tensor product of the Hilbert spaces.  I doubt that
it is.  

To define the category-theoretic tensor product requires
the notion of bimorphism.  For categories whose objects
are sets with additional structure, and whose morphisms are 
structure-preserving mappings we can define a bimorphism of $A,B$
as function $\phi: A \times B \rightarrow T$, where $T$ is another
object in the category, and $\phi$ has the property that for every
$a \in A, \phi_a: B \rightarrow T$ defined via $\phi_a (b) = 
\phi(a,b)$ is a morphism, and similarly with the roles of $A,B$ 
reversed.  In the category of vector spaces, for example, it is just
a bilinear map.  

\begin{definition}
The tensor product $A \otimes B$ is
a pair $(T, \tau)$, where $T$ is another object in the category 
(also often called the tensor product) and 
$\tau: A \times B \rightarrow T$ is a bimorphism, and any bimorphism
from $A \times B$ factors through $T$ in a unique way, and $T$ is
minimal among objects for which such a $\tau$ exists.
\end{definition}
To say $\tau$ factors through $T$ in a unique way is just to say that
for any bimorphism $\beta: A \otimes B \rightarrow V$, there is a
unique $\phi: T \rightarrow V$ such that $\beta$ is $\tau$
followed by $\phi$.  Minimality in a set means not a subobject of any
object in the set.  Probably the uniqueness of the factorization is
therefore redundant.

There is an ``operational'' motivation of this construction when it 
is applied to categories like effect algebras, operation algebras, 
etc...:  it implements the notion that the two structures being combined
appear as potentially ``independent'' subsystems of the larger system,
in a fairly strong sense that one can do any operation (or get any outcome)
on one subsystem while still having available the full panoply of operations
(outcomes) on the other.

The category-theoretic tensor product
of ordered linear spaces (vector spaces with distinguished regular
cones) is not well defined:  more structure is needed.  More precisely,
while various constructions having the universal property (all 
bimorphisms factor through them)  can be made,
there is not a unique minimal one.

For a variety of operational structures one might use to describe
quantum mechanical statics, including test spaces, orthoalgebras, and
effect algebras, the tensor product
is {\em not} the corresponding operational
structure for the tensor product of Hilbert spaces.  This could
indicate that the structure describing statics requires more
specialized axioms, still consistent with quantum mechanics, and then
the tensor product in this new category, call it $\cz$, will come out
right in the Hilbert space case.  It could also be that the difficulty
is the static nature of the categories.  Indeed, the
category-theoretic tensor product of test spaces or effect algebras
includes measurements whose performance would seem to involve
dynamical aspects.  These are measurements describable as the
performance of a measurement $M$ on system $A$, followed by the
performance of a measurement $M_\alpha$, on $B$, where which
measurement $M_\alpha$ is performed is conditional on the outcome
$\alpha$ of the $A$-measurement.  The 
the tensor product of effect algebras must contain all product
outcomes, and it can be characterized as the effect algebra
``generated'' by requiring that it contain all the ``1-LOCC'' (local
operations with one round, in either direction, of classical
communication) measurements just described.  Fuchs'
(\citeyear{Fuchs2001a}) ``Gleason-like theorem for product
measurements'' effectively does this construction for the
case of Hilbert effect algebras.  It is fairly elementary
to show that the tensor product of EA's 
can also be characterized as the minimal
``influence-free'' effect algebra containing all product measurements
(i.e. in which we can do all pairs of measurements one on $A$, one on
$B$, with no communication).  Freedom from influence of $B$ on $A$
means that for all states on the object, the probabilities of the
outcomes of an $A$ measurement, performed together with an independent
$B$ measurement, cannot be affected by the choice of measurement on
$B$.  Influence freedom means freedom from influence in both
directions.  Both of these characterizations
provide strong operational motivation for
the category-theoretic tensor product in this situation.
Each is
easily established starting from the other, 
and a similar construction of a ``directed'' product, in which 1-LOCC
operations are allowed in one direction only, rules out ``influence''
in the direction opposite the communication.  These things are also
true, and were in fact first established for, test spaces 
\citep{Foulis81a} and
orthoalgebras \cite{MKBennett93a}.

The difficulty, in the quantum case, is that the tensor product of
orthoalgebras or effect algebras, while it must contain measurements
of effects that are tensor products of Alice and Bob effects, and,
through addition of effects, all {\em separable} effects, does not
contain ``entangled'' Alice-Bob effects.  The separable effects span
the same vector space $B(\C^d \otimes \C^d) \cong H_{d^2}$ of $d^2
\times d^2$ Hermitian matrices (where A,B both have dimension $d$) as
the full set of effects on $\C^d \otimes \C^d$, but they are the
interval $[0,I]$ in the separable cone, not the interval $[0,I]$ in
the positive semidefinite cone.  Consequently the available states,
while they must be linear functionals of the form $A \mapsto \tr AX$
for $d^2 \times d^2$ Hermitian $X$, are the normalized members of the
separable cone's dual, rather than of the positive semidefinite cone's
dual, so $X$ in the functional $A \mapsto \tr AX$ is not necessarily
positive semidefinite.  The separable cone being properly contained in
the positive semidefinite one, its dual properly contains the positive
semidefinite one's dual, so that not only are we restricted to fewer
possible measurements, but their statistics---even those of
independent $A,B$ measurements---can be different from the quantum
ones (although all quantum states are also possible states).  Stated
in more quantum information-theoretic terms: some nonpositive
operators $X$ are nonpositive in ways that only show up as negative
probabilities or nonadditivity when we consider entangled
measurements: since in the effect-algebra or orthoalgebra tensor
product we don't have entangled measurements available to ``directly
detect'' this nonpositivity, these are admissible states on these
tensor products.  Indeed, as observed in \cite{Wilce92a}, they are
isomorphic to the Choi matrices (block matrices whose blocks $M_{i,j}$ 
are $T(\outerp{i}{j})$) of positive, but not necessarily
completely positive, maps $T$ 
(although the normalization condition (trace-preservation) 
appropriate for such maps is different from the (unit trace)
normalization condition appropriate
for states).  Of course, the nonpositivity of the operator
can be ``indirectly detected'' by tomography using separable effects,
since these effects span the space of Hermitian operators.  

One obvious solution to the problem would be to introduce axioms that
would prohibit this divergence between the existence of entangled
states and nonexistence of entangled measurements.  Mathematically,
this divergence reflects the important fact that the positive
semidefinite versus separable effect algebras on $\C^d \otimes \C^d$
are differentiated by the properties of the corresponding cones: the
former, but not the latter, being self-dual.  Self-duality is a 
natural and powerful mathematical requirement on cones, but
a very strong, and arguably not operationally motivated, one.
Self-duality is an important part of the essence of quantum
mechanics, so we should strive hard to understand its operational
motivation and implications.  
The cones for classical effect algebras can also be
self-dual: e.g. the algebra of fuzzy sets of $d$ objects.  
An axiom related to self-duality, violated by the tensor product of
Hilbert effect algebras,  is the ``purity is testability axiom.''  
We
develop some concepts before formulating it.
\begin{definition} An {\em effect-algebra theory} is a pair 
$\langle \ce, \Upsilon \rangle$ where $\ce$ is an effect algebra,
$\Upsilon$ a convex set of states on that effect algebra.
An effect $t$ {\em passes} a state $\omega$ if $\omega(t) = 1$.  An
effect $t$ is a {\em test} for $\omega$ in theory $\langle \ce,
\upsilon \rangle$ if $t$ passes $\omega \in \Upsilon$ and for no state
$\sigma \ne \omega, \sigma \in \Upsilon$, does $t$ pass $\sigma$.  A
state $\omega \in \Omega$ is {\em testable} in $\langle \ce, \Omega \rangle$
if a test for it exists in $\ce$.
\end{definition}

Note that $\Upsilon$ may be smaller than $\Omega(\ce)$, the set
of all possible states on $\ce$. 
We now assume effect algebras are convex.
If two tests pass $\omega$, so does any mixture of those 
tests.
Let $t$ be a test for $\omega$, then for $\sigma \ne \omega$,
$(\lambda \omega + (1 - \lambda) \sigma)(t) = 
\lambda \omega(t) + (1 - \lambda) \sigma(t) < 1$, i.e. $t$ cannot
test any mixture of $\omega$ with something else.  
Although a test thus tests a unique state, it is
not necessarily the case that a testable state has a unique test.
Let $t$ test $\omega$;  suppose $\omega = \lambda \sigma
+ (1 - \lambda) \tau$.  Then $1= \omega(t) = \lambda \sigma(t)
+ (1 - \lambda) \tau(t)$.  This implies that $\sigma(t) = \tau(t) = 1$,
hence by the fact that $t$ tests $\omega$, $\sigma = \tau = \omega$.
In other words, only pure (extremal) states can be testable.
We will be interested in 
{\bf Axiom 1:} {\em all} pure states are
testable. 
To study the consequences of this axiom, we introduce a basic 
notion in convex sets.
\begin{definition}
A {\em face} of a convex set $C$ is an $F \subseteq C$ such that
for every point $p \in F$, all points in terms of which $p$ can
be written as a convex combination are also in $F$.  In other 
words, for $\lambda_i \ge 0, \sum_i \lambda_i = 1$,  
$
\sum_i \lambda_i x_i \in F \Rightarrow (\forall i,~ x_i \in F)\;.
$
\end{definition}
Thus a face of $C$ is the intersection of the affine plane it generates with
$C$.  The set of faces, ordered by set inclusion, forms a lattice.
This lattice characterizes the convex set.
(up to affine isomorphism, 
which is the proper notion of isomorphism for convex sets since affine
transformations $y \mapsto Ay + b$ 
commute with convex combination).

\begin{proposition}
The theory 
$\langle \ce(\C^d) \otimes \ce(\C^d), \Upsilon
\rangle$ violates Axiom 1 unless $\Upsilon$ is contained
in the set of separable states.  In particular, 
$\langle \ce(\C^d) \otimes \ce(\C^d), \Omega(\ce(\C^d) \otimes \ce(\C^d)) \rangle$
violates it.
\end{proposition}
\begin{proof}
The proof proceeds by showing that the only states testable in
$\ce(\C^d) \otimes \ce(\C^d)$ are pure product states. Then if Axiom 1
is satisfied, the extremal states of $\Upsilon$ are product states, so
$\Upsilon$ is a face of the convex set of separable states.  Let $\tr
X = 1$ and $\bra{\chi}\bra{\psi} X \ket{\psi} \ket{\chi} \ge 0$ for
all product states $\kett{\phi}{\chi}$, so that $A \mapsto \tr AX$ is
a state.  Testability means there is a separable $A$ with trace
between zero and one (separable effect) such that:
$1 = \tr A X \;.$
The first requirement on $A$ says that 
$A = \sum_i \lambda_i \ket{\chi_i} \ket{\psi_i}
\bra{\psi_i} \bra{\chi_i}$ 
(for $\lambda_i > 0,
\sum_i \lambda_i \le 1,  
\ket{\chi_i}, \ket{\psi_i}$ normalized).
Thus $\tr AX = 1$ becomes 
$\sum_i \lambda_i 
\bra{\chi_i}\bra{\psi_i} X \ket{\psi_i} \ket{\chi_i} = 1$, which 
can only hold if one of the $\lambda_i =1$,
and for that $i$,  
$\bra{\chi_}\bra{\psi_i} X \ket{\psi_i} \ket{\chi_i}=1$.
Then (dropping the subscript)
$
X = \projj{\chi}{\phi} + X^{\pi,\perp} + X^{\perp,\pi} + X^{\perp,\perp}\;.
$
This is a resolution of $X$ into components in four subspaces of the
space of operators on $\C^d \otimes C^d$:  the space $\pi,\pi$ of operators on 
the one-dimensional Hilbert space $\pi$ spanned by the pure product
state, the space $\pi,\perp$ of operators taking $\pi$ 
to $\pi^\perp$, the space $\perp, \pi$ going the other way, 
and the space $\perp, \perp$ of operators on $\pi^\perp$.  The middle
two pieces are manifestly traceless, so the last one must be traceless
for $\tr X = 1$ to hold.  However, 
$\tr X^{\perp,\perp} = \sum_{ij}\braa{i}{j} X \kett{j}{i}$ 
in a product basis $\kett{i}{j}$ for $\perp$.
Each $\braa{i}{j} X \kett{j}{i}$
must be positive since $\tr X^{\perp,\perp} A = \tr X A$ for $A \in 
\perp,\perp$.  So for $X^{\perp,\perp}$ to be traceless, they must 
all be zero, and $X = \projj{\chi}{\phi}$ plus possibly some 
traceless stuff which does not affect the induced state.
\end{proof}

Note that we {\em can} have a theory on $\ce(\C^d) \otimes \ce(\C^d)$
satisfying the axiom of testability, but {\em only} if the state space is contained
in the dual of the cone generated by the effect algebra.  This suggests
that the axiom, if required of the full state space $\Omega$ of an 
effect algebra, is pushing us towards the idea that the cone be self-dual.

Testability is very natural, and has a long
history in quantum logic (e.g. \cite{Mielnik69a} and
probably Ludwig (\citeyear{Ludwig83a, Ludwig85a})).
Theories which are the full state spaces
of linear effect algebras that are initial intervals 
in self-dual cones satisfy it.
This axiom makes contact with the ``property lattice'' quantum logics
of \cite{Jauch68a} and \cite{Piron76a}.  
(See \citet[pp. 220--221]{Valckenborgh2000a}). 
It is also related to R\"{u}ttiman's 
\citet{Ruttiman81a} notion of ``detectable property.''
Jauch and Piron's
notion of property roughly corresponds to effects (or the analogues in
other quantum structures, since most of their work was done before
effect algebras were formalized in the quantum logic community) $e$
which can have probability one in (``pass'') some states.  
Those states are said to ``possess
the property $[e]$''.  Properties are equivalence
classes of effects that pass the same set of states.
They construct a lattice of properties for an
empirical theory (set of states on some quantum structure).  

Axiom 1 relates the
lattice of faces of a convex set of states on an effect algebra 
to the property lattice of that
theory.  
The extremal states are minimal in the
face lattice;  the axiom says there are ``minimal properties''
possessed by those states: minimal in the sense that no other state
posesses them.  I am not certain if this is minimality in the sense
of Piron's property lattice, but it seems likely
(perhaps under mild conditions).
A generalization of Axiom 1 asserts, for each face of the 
state-set, the existence of a ``property'' of being in that
that face (an
effect passing the states of that face and no others).  
A similar axiom of \cite{Araki80a}
concerns ``filters'' for higher dimensional 
faces, 
but this also involves ``projection postulate-like''
dynamics associated with the filtering.  Araki also uses, as an
assumption, the symmetry or ``reciprocity'' rule, 
satisfied in the quantum-mechanical
case, that can be formulated once a correspondence $\chi 
\leftrightarrow e_\chi$ between
extreme states $\chi$ and tests $e_\chi$ for them has been set up.
Reciprocity requires that
$
\chi(e_\phi) = \phi(e_\chi)\;.
$
It is not clear to me whether the extreme states $\rightarrow$
effects correspondence must be one-to-one
instead of many-to-one
in order to be able to formulate the axiom, or whether one-to-oneness
might be a consequence of it.
(Faces play an important role in Ludwig's work as well, as do
statements reminiscent of Axiom 1, so Ludwig's
argument may turn out to be similar.)

Araki credits Haag for emphasizing to him the importance of the
reciprocity axiom.  In the second edition of his book,
\cite{Haag96a} includes a informal discussion of the foundations of quantum
mechanics based on the convex cones framework.  He, too, uses
Axiom 1, and a generalization associating faces of the state
space (one-to-one!) with 
``propositions.''  These ``propositions'' are 
effects passing precisely the states
of the face, and minimal among such effects in the sense of a 
probabilistic ordering of effects
$
e_1 \le e_2 := \forall \omega \in \Upsilon ~\omega(e_1) \le \omega(e_2)\;.
$
This is a different strategy from the Jauch-Piron equivalence class
one for getting uniqueness of the effect associated to a face, but
it is closely related to it.
Jauch and Piron were trying to get by with less reference to probabilities.
Haag also uses the reciprocity axiom, which he argues imposes 
self-duality.\footnote{Haag uses uses the notion of 
self-polarity, but for our type of cone, this is the same as self-duality.  
The polar of a convex body $C$ is the set of linear
functionals $L$ such that $L(x) \le 1$ for all $x \in C$; the polar of
a cone is the negative of the dual cone, since whenever $L(x)$ is 
positive, $L(x')$ is greater than $1$ for $x'$ a large enough positive
multiple of $x$.  Since the negative of a cone is isomorphic to that cone,
a self-polar cone is self-dual.}

Haag also gives some operational motivation for an additional assumption,
that of homogeneity of the cone.  This says that the 
automorphism group of the cone acts transitively on its interior.  
(For any pair $x,y$ of interior points, there is 
an automorphism taking $x$ to $y$.)  Interpret
cone automorphisms as conditional dynamics; then 
homogeneity, at least for self-dual cones,
means that any state 
is reachable from any other by dynamics conditional on
some measurement outcome.  This is not self-evident
but seems natural.  If you can't prepare any 
state starting from any other state, with a nonzero probability
of success, the state space might ``fall apart'' into pieces not reachable
one from the other (orbits of the automorphism group).  
Or while some pieces might still be
reachable from all others, going the other way might not be possible:
there would be intrinsically irreversible dynamics, 
even conditionally.  A more detailed study of 
operational theories whose effects are naturally 
represented in a non-homogeneous
cone, or whose state-space generates one, would
be desirable (either with or without self-duality).
The ``falling apart'' into orbits of the automorphism group
may be acceptable in a theory of a perspective
involving radical limitations on our ability to prepare states:
going from one orbit to another might require a more powerful agent
than the one whose perspective is being considered, but the consequences
of such an agent's actions might be observable by the less powerful
agent.  Entanglement is such a situation:  the
perspective of the set of local agents, even with the power to communicate
classically, allows for pairs of states with
different statistics for observables implementable by local
actions and classical communication (LOCC), such that it is 
impossible, even
conditional on a measurement outcome, to prepare one starting from
the other via LOCC \cite{Dur2000a}.  
The LOCC perspective of the local agents is not 
usually taken as a ``subsystem'' in quantum mechanics, so these sorts
of perspectives can there be taken as derivative rather than fundamental;
but perhaps in other situations nonhomogeneous perspectives
could be more fundamental.

In finite dimensions, as Haag points out, homogeneous self-polar cones
are known (e.g. \citep{Vinberg65a}) to be isomorphic to direct products of 
the cones whose faces
are the subspaces of complex, quaternionic, or real Hilbert spaces.
(Extensions of these results to infinite dimensions are obtained in
\cite{Connes74a}.)  The factors in the direct product can be thought of as 
``superselection sectors;''  classical theory would be recovered 
when the superselection sectors
are all one-dimensional (at least in the complex and 
real cases).   
\cite{Araki80a} obtains a similar theorem except
the effects get represented as elements of a finite dimensional
Jordan algebra factor.  
These are isomorphic to to $n \times n$ 
Hermitian matrices over $\R, \C$, or the quaternions 
$\H$, or a couple of exceptional
cases (spin factors and $3 \times 3$ Hermitian matrices over the
Cayley numbers).  
He also gives arguments for picking
the complex case, based on the properties of composition 
of subsystems in the various cases.
Araki's argument is that ``independence'' of the
subsystems should be expressed by $\text{ dim } V = 
(\text{ dim } V_1)(\text{ dim } V_2)$ for the algebras.
But, ``essentially because the tensor product of two skew-Hermitian 
operators is Hermitian'', we have 
$\text{ dim } V >
(\text{ dim } V_1)(\text{ dim } V_2)$ except in trivial cases, when we
take the $V$'s to be the algebras of Hermitian matrices over
real Hilbert spaces $H_1$, $H_2$, and their tensor product.
(A related requirement plays a similar role in \cite{Hardy2001a,Hardy2001b}.)
For $\Q$ there is not even a quaternion-linear tensor product.
The bottom line is that ``the complex field has the most pleasant feature
that the linear span of the state space of the combined system is
a tensor product of [the state spaces of the] individual ones.''
There are probably important operational and
information-theoretic distinctions between the cases which merit closer
study.
In the real case, the key point is that in contrast to 
the complex case, states on the ``natural''
real composite system are not determined by the expectation values of
local observables.

Like homogeneity, self-duality and reciprocity 
may be related to the coordination of perspectives into an overall structure.
In a ``spin-network'' type of theory, the edges of a graph are labelled
with representations of a Lie or quantum group ($\fsu(2)$, 
for spin networks), which are Hilbert spaces.  The vertices are associated
to ``intertwiners'' between those representations.  A state might 
be associated with, say, a partition of the graph by a hypersurface
cutting it into two parts, ``observer'' and ``observed.''  If the hypersurface
has two disconnected parts, the associated Hilbert space 
will be the tensor product of the ones associated with 
the parts;  otherwise, the representation is made out of the representations
labelling the cut edges, in a way determined by the intertwinings at the
vertices between them.  One has
the same Hilbert space whichever piece one takes as ``observer''
vs. ``observed.''  However, it is likely that the role-reversal
between observer and observed corresponds to dualization, and the
result that both correspond to the same Hilbert space will only 
hold in theories in which the structure describing a 
given perspective---here, the Hilbert space associated with the 
surface---is self-dual.  To attempt to actually show something
like this would involve a project of trying to 
construct ``relational'' theories like the Crane-Rovelli-Smolin
theories, but with other empirical theories playing
the role of Hilbert spaces and algebras of observables on them.
A simple first example might be  
``topological classical field theories,'' if these can consistently
be defined.  
In these general ``pluralistic structures'' 
coordinating perspectives, one 
might hope to find a role for self-duality and the reciprocity
axiom, and perhaps homogeneity as well.  For the different empirical
structures associated with  different surfaces to relate to each other
in a ``nice'' way, it might be necessary that the structures be
defined on self-dual cones or exhibit reciprocity.  
Another suggestion that bears more detailed investigation, perhaps
also in the ``relational'' 
context since there time is sometimes taken as emergent, 
is due to Haag, who says,
``[reciprocity] expresses a symmetry between ``state preparing 
instruments'' and ``analyzing instruments'' and is thus related 
to time-reversal invariance.''

\section{Tasks and axioms: toward the marriage of quantum information science and operational quantum logic}
\label{sec: applying operational logic}

QIP emphasizes how the conceptual peculiarities of quantum
mechanics allow us to perform tasks not classically possible.  This
suggests we these formulate tasks, or the 
associated concepts, in ways general enough to try to  
characterize different operational theories by whether or not these
tasks can be performed in them, or by the presence or absence of
conceptual phenomena such as:  superposition, complementarity, 
entanglement, 
information-disturbance tradeoffs, restrictions on cloning or 
broadcasting, nonuniqueness of the expression of states
as convex combinations of extremal quantum states (versus the 
uniqueness classically), and so forth.  
An outstanding example involves cryptographic tasks
\citep{Fuchs2001a,CBH2002a}.
But even before the upsurge of interest
in quantum information science, conceptual peculiarities 
like superposition \citep{MKBennett90a} and 
nonunique extremal
decomposition 
\citep{Beltrametti93a}
were being
generalized and studied in empirical/operational quantum logic.  

Assumptions and tasks involving computation should also be investigated;
In particular, it would
be interesting to establish linkages between complementarity, 
or superposition,
and computational speedup in a general setting.
Or some conjunction of 
properties, such as no instantaneous communication between subsystems, 
common to quantum and classical mechanics, might be seen to imply no 
exponential speedup of brute-force search in a general setting.
I claimed above that key aspects of using an operational point of
view in foundational questions were understanding notions of subystems
and system combination, and understanding dynamics.
For information-processing or computation, both of these
issues are of the utmost importance.
Since the environment which induces noise in a system or the apparatus
used by an information-processing agent must be considered together
with the system, a notion of composite system is needed.  And notions
of composition of systems or of dynamics 
are basic to computational complexity, where the
question may be how many bits or qubits are needed, as a function of
the size of an instance of a problem (number of bits needed to write
down an integer to be factored, say) to solve that instance.  
The very notion of Turing computability is based on a factorization of
the computer's state space (as a Cartesian product of bits, or of some
higher-arity systems), in terms of which a ``locality'' constraint can
be imposed.  The constraint is, roughly, that only a few of these
subsystems can interact in one ``time-step.''  The analogous quantum
constraint allows only a few qubits to interact at a time.  In general
operational models, some notion of composition of systems, such as a
tensor product, together with a theory describing what dynamics can be
implemented on a subsystem, could allow for generalized 
circuit or Turing-machine
models.  Another way of obtaining a notion of resources
is to specify a set of dynamical evolutions to
which we ascribe unit cost, and a set of
measurements viewed as computationally easy.  More generally, we might
specify a cost function on evolutions and measurements.
A formal treatment
will require us to say how we
interface the given operational model with
``classical'' computation.  We could specify a set of
measurements-with-conditional-dynamics (``instruments'') viewed as
taking unit computational time, and allow the conditioning of further
dynamics and measurement on the results of the measurement in
question.  Subtleties could arise in counting the computational cost
of the classical manipulations required by 
such conditioning.
Counting one elementary operation in some chosen classical computational
model as costing the same as
one in the general operational model is one reasonable
approach (at least if the
general model can simulate classical computation polynomially).  More
simply, perform the algorithm in the general operational setting by
evolution without explicit measurement and classical control, and
specify a ``standard'' measurement to be performed at the end (and
a standard procedure for mapping
the measurement result to the set of possible values of the function
being computed).  In non-query models, it is
important that not just any measurement be allowed at the end, since
if the dynamics consists of all effect-algebra endomorphisms, say, any
computation can be done by making one measurement.

\section{Conclusion}

In this paper, I have promoted a particular project for harnessing the
concepts of quantum information science to the task of illuminating
quantum foundations.  This project is to generalize tasks and concepts
of information science beyond the classical and the quantum, to
abstract and mathematically natural frameworks that have been
developed for representing empirical theories; and to use these tasks
and concepts to develop axioms for such theories, having intuitively
graspable, perhaps even practical, meaning, or to develop a better
understanding for the operational meaning of existing axioms.  The
main original technical contributions are Theorem
\ref{theorem: main} showing that any phenomenological theory naturally
gives rise to a weak effect algebra, which is essentially
the image of the propositional
logic of statements about measurement outcomes under identification of
probabilistically
equivalent outcomes, and the introduction of the notions of operation
algebra and weak operation algebra.  These results and concepts
are likely  closely
related to other work in operational quantum logic and the convex
approach; I think they provide an appropriate framework for the
project.

Within the scope of this project, I have emphasized what I think
will be key aspects:
\begin{itemize}
\item
A ``perspectival, operational'' approach to describing empirical theories, 
taking the probabilities of outcomes of operations
an agent may do on the system as primary, and stressing that
the 
structure of an empirical theory depends on the agent
doing the operations as well as
on the subystem the operations are done on.
\item
The structures of effect algebras and weak effect 
algebras, test spaces, and proposition lattices for 
observations, as well
frameworks of ``operation algebras'' and 
``weak operation algebras'' introduced here to encompass 
both dynamics
and observables.
\item
A justification of  
weak effect and operation algebras through relations of
``probabilistic equivalence,'' and ``sequential probabilistic
equivalence,'' as natural representations
of very general classes of phenomenological theories.
Gleason-type theorems take on a fresh aspect from this point of view.
\item
Convexity, and the 
resulting representations it makes possible
in ordered linear spaces (real vector spaces with distinguished
regular cones), and
various mathematically natural axioms it suggests, such as 
homogeneity and self-duality.
\item
The significance of other natural operational desiderata, such as 
the idea that anything implementable via interaction with an independent
ancilla should be considered an operation, or the idea that ``evolve and
then measure'' should be considered a kind of measurement.
\item
The importance of 
attempts, like the Rovelli-Smolin ``relational quantum 
mechanics,'' 
topological quantum field theories, spin networks, and
``spacetime foams,''
to integrate agents' perspectives into a coherent
whole, as special relativity does with its reference frames.
The use of ``integrability of perspectives into a coherent whole,''
as a possible source of axioms about the nature of perspectives
(self-duality or homogeneity of the cones used to represent them?), 
how they combine (via tensor products or some other rule?), and
so forth.
\end{itemize}
  
\section*{Acknowledgments}
Discussions over the years with Carlton Caves, Dave Foulis, 
Chris Fuchs, Leonid Gurvits, Lucien Hardy, Richard Jozsa,
Manny Knill, 
Eric Rains, R\"udiger Schack, Ben Schumacher, and Alex Wilce, 
among others, have influenced
my thoughts on these matters.  

\bibliographystyle{elsart-harv.bst}

\end{document}